\documentclass[12pt,document,nofootinbib,superscriptaddress,onecolumn,preprintnumbers,balancelastpage]{revtex4}
\pdfoutput=1
\usepackage{latexsym}
\usepackage{amssymb}
\usepackage{epsfig,amsmath,graphics}
\usepackage{listings}
\usepackage{color}
\usepackage{dcolumn}
\usepackage{slashed}
\usepackage{cancel}
\usepackage{comment,latexsym} 
\usepackage{bm}
\usepackage{verbatim}
\usepackage{tabularx}
\usepackage{dcolumn}
\definecolor{Mahogany}{rgb}{0.62,0.24,0.15}
\definecolor{colorLink}{rgb}{0.7,0,0}
\definecolor{colorCite}{rgb}{0,.7,0}
\definecolor{colorURL}{rgb}{0,0,0.7}
\definecolor{colorTC}{rgb}{.2,.7,.2}
\definecolor{colorDP}{rgb}{.7,.7,.2}
\usepackage[colorlinks=true,linktocpage=true,linkcolor=black,citecolor=colorCite,urlcolor=colorURL]{hyperref}
\usepackage{setspace}
\usepackage{xspace}
\usepackage{multirow}
\usepackage{titlesec}
\usepackage[bbgreekl]{mathbbol}
\usepackage{bm}

\newcommand{\OO}{\mathcal{O}}

\newcommand{\ZZ}{\mathbb{Z}_2}

\newcommand{\Li}{{\rm Li}}
\newcommand{\wt}{\widetilde}
\renewcommand{\Re}{{\rm \,Re\!}}
\renewcommand{\Im}{{\rm \,Im}}
\newcommand{\abs}[1]{\left|\, #1 \,\right|}

\newcommand{\f}{$f$-}

\def\be{\begin{equation}}
\def\ee{\end{equation}}
\newcommand{\beq}{\begin{equation}}
\newcommand{\eeq}{\end{equation}}
\newcommand{\eref}[1]{Eq.~(\ref{#1})}

\newcommand{\lsim}{\!\mathrel{\hbox{\rlap{\lower.55ex \hbox{$\sim$}} \kern-.34em \raise.4ex \hbox{$<$}}}}
\newcommand{\gsim}{\!\mathrel{\hbox{\rlap{\lower.55ex \hbox{$\sim$}} \kern-.34em \raise.4ex \hbox{$>$}}}}
\newcommand{\vev}[1]{ \left\langle {#1} \right\rangle }
\newcommand{\GeV}{{\text{ GeV}}}

\newcommand{\eg}{\emph{e.g.}}

\expandafter\def\expandafter\normalsize\expandafter{%
    \normalsize
    \setlength\abovedisplayskip{8pt}
    \setlength\belowdisplayskip{8pt}
    \setlength\abovedisplayshortskip{8pt}
    \setlength\belowdisplayshortskip{8pt}
}

\usepackage{feynmp-auto}
\titleformat{\section}{\center\normalfont\fontsize{14}{15}\bfseries}{\thesection.}{1em}{}
\titleformat{\subsubsection}{\center\normalfont\fontsize{12}{15}}{\thesubsubsection.}{1em}{}

\begin{document}
$\quad$
\vskip 60 pt

\title{Folded Supersymmetry with a Twist} 

\author{Timothy Cohen}
\affiliation{
Department of Physics, Princeton University, Princeton, NJ 08544
\vspace{-4pt}
}
\affiliation{
School of Natural Sciences, Institute for Advanced Study, Princeton, NJ 08540
\vspace{-4pt}
}
\affiliation{
Institute of Theoretical Science, University of Oregon, Eugene, OR 97403
\vspace{-4pt}
}

\author{Nathaniel Craig}
\affiliation{
Department of Physics, University of California, Santa Barbara, CA 93106
\vspace{-4pt}
}

\author{Hou Keong Lou}
\affiliation{
Department of Physics, Princeton University, Princeton, NJ 08544
\vspace{-4pt}
}

\author{David Pinner$^{\,}$}
\affiliation{
Princeton Center for Theoretical Science, \\
\vspace{-8pt}
Princeton University, Princeton, NJ 08544 }

\begin{abstract}
\vskip 1 pt
\begin{center}
{\bf Abstract}
\end{center}
\vskip -30 pt
$\quad$
\begin{spacing}{1.05}\noindent
Folded supersymmetry ($f$-SUSY) stabilizes the weak scale against radiative corrections from the top sector via scalar partners whose gauge quantum numbers differ from their Standard Model counterparts. This non-trivial pairing of states can be realized in extra-dimensional theories with appropriate supersymmetry-breaking boundary conditions. We present a class of calculable $f$-SUSY models that are parametrized by a non-trivial twist in 5D boundary conditions and can accommodate the observed Higgs mass and couplings.  Although the distinctive phenomenology associated with the novel folded states should provide strong evidence for this mechanism, the most stringent constraints are currently placed by conventional supersymmetry searches.  These models remain minimally fine-tuned in light of LHC8 data and provide a range of both standard and exotic signatures accessible at LHC13.
\end{spacing}
\end{abstract}

\maketitle
\newpage
\begin{spacing}{1.3}
\pagebreak
%
%

\section{Introduction}
\label{sec:Intro}
The canonical solutions to the naturalness problem of the Standard Model are based on supersymmetry (SUSY) or compositeness.  Both of these approaches lead to ``top partner" states that carry the Standard Model gauge quantum numbers of the top quark and couple to the Higgs boson with a strength set by the top Yukawa.  The QCD charge of the top partners implies a large production cross section at the LHC.  Searches for these states have been performed using data from LHC8, and the null results can be interpreted as constraining naturalness in the context of these models (see \eg~\cite{Craig:2013cxa, Bechtle:2015nta} for reviews on the status of SUSY). In particular, the non-observation of partner states gives rise to a ``little'' hierarchy problem -- percent-level cancellations between threshold corrections are required to reconcile the weak scale with experimental bounds on the new states predicted by supersymmetry or compositeness. 

However, as was first pointed out in the seminal papers~\cite{Chacko:2005pe, Burdman:2006tz}, it is logically consistent for the top partner states to be color-neutral.  The resulting signatures can differ wildly from those traditionally associated with naturalness.  The production cross-section of the un-colored top partners is significantly reduced, with correspondingly weaker limits at the LHC~\cite{Burdman:2014zta}.  Lighter top partners are thus consistent with data, ameliorating the tension of the little hierarchy problem.  There are two broad classes of these so-called ``neutral natural" models:  Twin Higgs models \cite{Chacko:2005pe, Chacko:2005vw, Chacko:2005un, Chang:2006ra, Batra:2008jy, Craig:2013fga, Craig:2015pha} (or their generalization into the Orbifold Higgs~\cite{Craig:2014roa}) rely on a shift symmetry to protect the Higgs potential, whereas models of Folded SUSY (\f SUSY) are supersymmetric at short distances~\cite{Burdman:2006tz}. In both cases, discrete symmetries enforce the equality of the Higgs couplings to the top quark and its un-colored partners. This results in the accidental partial restoration of the UV symmetry at low energies. 

For models with shift symmetries in the UV, \emph{e.g.}~Twin Higgs, the Higgs quartic must necessarily be generated at loop level (with the possible exception of terms arising from collective symmetry breaking~\cite{Chacko:2005vw, Cai:2008au}).  Since the Higgs mass parameter is generically one loop down from the scale of global symmetry breaking $f$, naturalness suggests that $v \sim f$, where $v$ is the Higgs vacuum expectation value (vev). Furthermore, misalignment of the Higgs boson from the SM vacuum leads to irreducible Higgs coupling deviations parametrized by $(v/f)^2$; experimental constraints derived from Higgs properties require $f\gtrsim 700 \GeV$~\cite{ATLAS-CONF-2014-010}, imposing a minimum tuning cost irrespective of the direct search reach for partner states.  In contrast, neutral natural models relying on supersymmetry, such as \f SUSY, can accommodate both a tree-level quartic and a Higgs boson aligned with the Standard Model vacuum. As such, top partner masses can be pushed as high as $\sim 500$ GeV with no attendant loss of naturalness.  
  
How does one construct a model of neutral naturalness that is consistent with SUSY at high energies?  Specifically, since the $\mathcal{N} = 1$ global supersymmetry commutes with the Standard Model gauge group, the low-energy pairing of colored and un-colored states in \f SUSY resembles hard SUSY breaking from a 4D perspective.  Clearly both the colored and un-colored states must come in full supermultiplets.  Then the desired low energy spectrum must be realized through SUSY-breaking.  This can be naturally realized in 5D theories compactified on $S_1/\ZZ$ with suitable boundary conditions.  Since SUSY breaking is realized non-locally in the extra dimension, \emph{i.e.}, the Scherk-Schwarz mechanism~\cite{Scherk:1978ta, Scherk:1979zr}, contributions to the Higgs potential are finite despite the inherently low cutoff scale typical of 5D models.   

Given this feature of calculability, it is interesting to undertake a careful analysis of electroweak symmetry breaking in \f SUSY.  In this paper we describe the relation between the allowed parameter space and weak scale fine-tuning.  We will begin by analyzing \f SUSY in its original incarnation, wherein the un-colored $f$-top partners are massless at tree-level while their colored counterparts have a large mass set by the compactification scale.  As will be shown below, the Higgs potential (given by an approximate two-loop calculation) does not appear to yield a non-zero vacuum expectation value for the Higgs boson.  This motivates deforming minimal \f SUSY in order to simultaneously achieve both electroweak symmetry breaking and a 125~GeV Standard Model-like Higgs boson.  One minimal extension is to allow for a non-trivial twist $\alpha \in[0,1/2]$ in the Scherk-Schwarz boundary conditions, leading to tree-level masses for the zero-mode $f$-scalars and thereby controlling a partial cancellation between top and \f top sector contributions to the Higgs potential. This additional parametric freedom provided by $\alpha$ allows electroweak symmetry breaking to be triggered at one loop without spoiling the attractive features of finiteness and calculability.

In the minimally tuned regions of parameter space, \f SUSY models with non-trivial twist give rise to a range of observable phenomenology.  Generic predictions include many of the exotic signatures associated with the original model of \f SUSY.  For example, the $f$-squarks with mass $|\alpha-1/2|/R$ behave as scalar squirks~\cite{Kang:2008ea} that annihilate to electroweak resonances~\cite{Burdman:2008ek, Burdman:2014zta}.  There are also interesting hidden-valley-like~\cite{Strassler:2006im, Han:2007ae} observables associated with the $f$-glueball states~\cite{Harnik:2008ax, Craig:2015pha, Curtin:2015fna}.  Additionally, these models may be probed indirectly through precision Higgs measurements~\cite{Craig:2013xia, Fan:2014axa}.

The precise phenomenology depends in detail on both the Scherk-Schwarz twist and on the geographic distribution of supersymmetric multiplets in the extra dimension.  For concreteness, we will focus on two specific models distinguished by the details of the Higgs sector.  In the first model, the Higgs bosons will be restricted to a four-dimensional brane. In this case, we will see that the dominant constraint arises from a combination of precision Higgs measurements and searches for the heavy Higgs bosons.  In the allowed region of parameter space, this model is tuned at the $\OO(0.2 \%)$ level.  This motivates studying a second model where the Higgs fields are allowed to propagate in the bulk.  Now both the Higgsinos and the heavy Higgs states are lifted using boundary conditions.  This alleviates the tension with Higgs measurements and allows for a reduction in fine-tuning to better than the $10\%$ level.  The dominant collider signature will come from gluino-squark associated production. Intriguingly, compressed spectra are a generic feature of these models, weakening limits on radius of the extra dimension.  In both the brane and bulk Higgs cases, discovery will proceed through the detection of MSSM-like states, followed by the spectacular observation of the $f$-squark sector.

The rest of this paper is organized as follows:  In Sec.~\ref{sec:SS} we review the generic features of 5D supersymmetric theories with Scherk-Schwarz SUSY breaking.  In Sec.~\ref{sec:fSUSY} we describe the top and gauge sectors of \f SUSY with arbitrary Scherk-Schwarz twist, along with the radiatively-generated Higgs potential.  In Secs.~\ref{sec:BraneModel} and~\ref{sec:BulkModel} we present in detail the parameter space and fine-tuning implications of two complete models in which the Higgses are alternately localized on the brane at $y=0$ or allowed to propagate in the bulk.  Finally, we conclude in Sec.~\ref{sec:Conc}. We reserve a variety of technical results for a series of appendices.

\section{Supersymmetry in 5D}
\label{sec:SS}
In this section we review some basic properties of theories with Scherk-Schwarz SUSY breaking.  In particular, we consider 5D theories compactified on $S_1 / \mathbb{Z}_2$ in which SUSY is broken by boundary conditions on the orbifold fixed points.  As the smallest spinor in 5D is Dirac (with 8 real components), when written in terms of 4D superfields \cite{ArkaniHamed:2001tb} the matter content must be equivalent to that of $\mathcal{N} = 2$ SUSY.  Furthermore, the interactions in the bulk must respect the full $SU(2)\times U(1)_R$ symmetry.  Just as in the non-supersymmetric case, the low-energy theory will not contain chiral fermions unless suitable boundary conditions are chosen to ``project out'' the zero-mode Dirac partners.  A Dirichlet (Neumann) boundary condition then corresponds to an odd (even) eigenvalue under reflection about $y = 0$ ($\pi\,R$), hereafter denoted $\mathcal{Z}$ ($\mathcal{Z}'$).  This may be accomplished in an $\mathcal{N} = 1$ SUSY-invariant way by choosing $\mathcal{Z}$ and $\mathcal{Z}'$ to act on complete 4D superfields.

\subsection{Supersymmetry Breaking by Boundary Conditions}

The action of the transformations $\mathcal{Z}, \mathcal{Z}'$ on superfields are
\begin{equation}
\mathcal{Z} \big[\Phi(y)\big] = Z\, \Phi(-y) \hspace{2cm} \mathcal{Z}' \big[\Phi(y-\pi\, R)\big] = Z'\, \Phi(-y + \pi \, R),
\label{eq:ZandZPrime}
\end{equation}
where $Z^2 = Z^{\prime 2} = 1$. The apparent $\mathcal{N} = 2$ SUSY is broken down to $\mathcal{N} = 1$ if the eigenvalues are different for each of the $\mathcal{N} = 1$ superfields in an $\mathcal{N} = 2$ multiplet.  In particular, we may choose $Z^{\prime} = U \Sigma_3 U^{-1}$, in which $\Sigma_3$ is $+1$ $(-1)$ on even (odd) $\mathcal{N} = 1$ superfields.  Here we take $U$ to be a $2 \times 2$ matrix acting on the $SU(2)_R$ doublets; it acts trivially on $SU(2)_R$ singlets.  More generally, the eigenvalues $\Sigma_3$ can be chosen to break additional symmetries, and likewise the rotation matrices can act on other symmetry multiplets~\cite{Barbieri:2001dm}, although we ignore this possibility for simplicity.  

We will work in the basis in which $Z$ is diagonal, $Z = \Sigma_3$.  If boundary conditions are chosen such that the same $\mathcal{N} = 1$ SUSY is respected by both of the fixed points, then $U = \mathbb{1}$ and $Z' = \Sigma_3$, corresponding to $[Z, Z'] = 0$. In this case, the low-energy theory after KK decomposition will also be $\mathcal{N} = 1$ supersymmetric.  In general, however, the fixed points may be taken to be invariant under inequivalent SUSY transformations.  Then we can parameterize this commutator by a single angle (ignoring possible phases), $0 \leq 2\, \pi \,\alpha \leq \pi$, which parameterizes the rotation between bases in which $Z, Z'$ are respectively diagonal.\footnote{If other symmetries were also broken with boundary conditions, the commutator would be parameterized by additional angles corresponding to the orientation of $Z, Z'$ in the space of these additional symmetries.} In the diagonal basis of $Z$, $Z'$ is then given by
\begin{equation}
Z' = e^{- \pi \,i\,\alpha \, \sigma_2}\, Z\, e^{\pi \,i\,\alpha \, \sigma_2}\ .
\end{equation}

For $\alpha = 0$, $Z$ and $Z'$ commute as above, and the theory is $\mathcal{N} = 1$ supersymmetric.  At $\alpha = 1/2$, these $\mathbb{Z}_2$ transformations are ``orthogonal'' in the sense that again $\big[Z,Z'\big] = 0$, but $Z' = -Z$. For general $0 < \alpha < 1/2$, $Z$ and $Z'$ do not commute; in the basis where $Z = \Sigma_3$, we have
\begin{equation}
\big[Z,Z'\big] = e^{2 \,\pi \,i \, \alpha\, \sigma_2} - e^{- 2\, \pi\, i\, \alpha\, \sigma_2} \ .
\end{equation}

\begin{figure}[t!]
\centering
\includegraphics[width=.65 \textwidth]{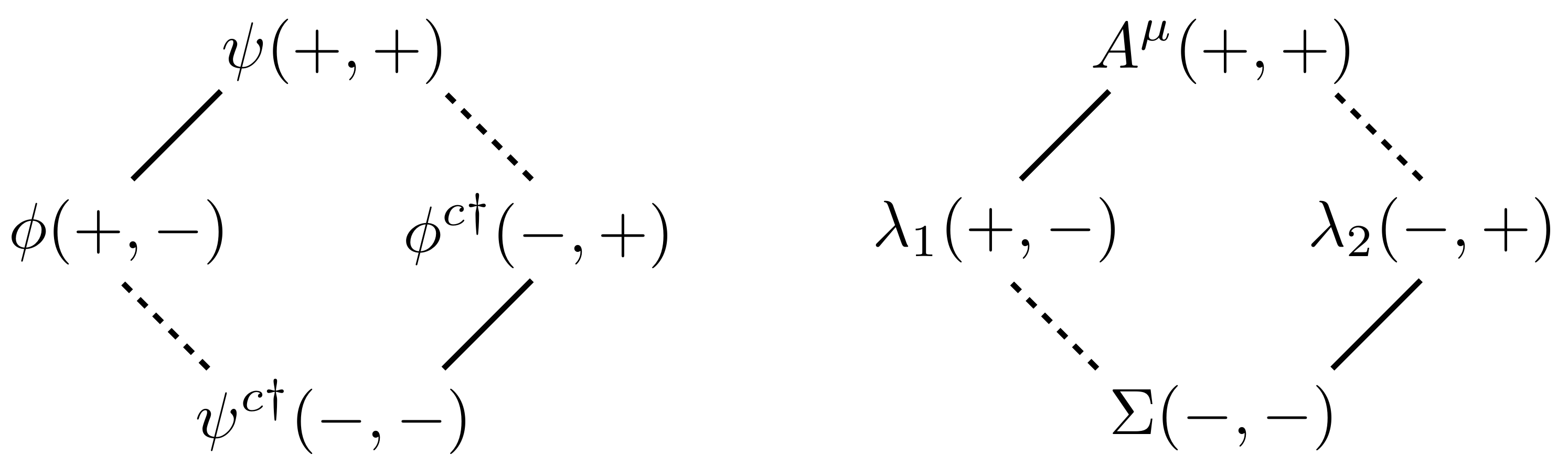}
\caption{Component field eigenvalues under reflections at $y = 0$ and $y = \pi\, R$, shown for a matter hypermultiplet (left) and a vector multiplet (right) at maximal twist.  Solid (dashed) lines represent component field groupings into 4D superfields at $y = 0$ ($\pi\, R$).}
\label{fig:Z2MaxTwist}
\end{figure}

As an example, consider a matter hypermultiplet, $\big(\Phi, \Phi^c\big)$, composed of a conjugate pair of 4D chiral superfields, $\Phi \sim (\phi, \psi)$ and $\Phi^c \sim (\phi^c, \psi^c)$.  Without loss of generality, we may take $\Phi$ and $\Phi^c$ to be eigenstates of $\mathcal{Z}$.  Then the eigenstates of $\mathcal{Z}'$ are, in general, inequivalent chiral superfields $\Phi' \sim (\phi', \psi)$ and $\Phi^{c \prime} \sim (\phi^{c \prime}, \psi^c)$, in which 
\begin{equation}
\left(\begin{array}{c} \phi' \\ \big(\phi^{c\, \prime}\big)^{\dag}\end{array}\right) 
= e^{\pi \,i\, \alpha \, \sigma_2} 
\left(\begin{array}{c} \phi \\ \big(\phi^{c}\big)^{\dag}\end{array}\right).
\end{equation}
At maximal twist ($\alpha = 1/2$), $\phi' = \phi^{c\dag}$ and $\phi^{c \,\prime} = -\phi^{\dag}$.  This situation is illustrated in Fig.~\ref{fig:Z2MaxTwist}.  After integrating out the extra dimension, the component fields may be KK decomposed, $\psi(x,y) \propto \sum_k f_k(y)\, \psi^{(k)} (x)$, with each mode having an apparent 4D mass $m^2_k\, f_k = \partial^2_y \, f_k$.    For example, given the boundary conditions in Fig.~\ref{fig:Z2MaxTwist}, the KK modes of each component field will be split in mass.  At low energies this is just a theory with softly-broken SUSY.  This approach to SUSY breaking is the Scherk-Schwarz mechanism; for explicit wave functions and arbitrary boundary conditions, see the Appendix.

Noting that the theory is locally $\mathcal{N} = 1$ invariant everywhere in the extra dimension, all SUSY-breaking effects must be global in nature.  In particular, all UV divergences, which correspond to contractible loops, must vanish supersymmetrically.  The SUSY-breaking contributions of these loops come from wrapping the virtual propagators around the extra dimension, so that they ``see'' the SUSY-breaking boundary conditions.  The space-like separation of these propagators gives an exponential suppression at high momenta, rendering all such loops finite~\cite{ArkaniHamed:2001mi, Contino:2001nj, Delgado:2001ex, Contino:2001gz, Kim:2001re}.  Therefore, although 5D models with bulk gauge fields are intrinsically non-renormalizable and necessarily come equipped with a UV cutoff $\Lambda$ (which in the models we will consider is not far above $1/R$), all SUSY-breaking effects will be calculable.  In particular, as we will see in the following sections, all corrections to the Higgs potential are explicitly finite.

Another generic feature of Scherk-Schwarz SUSY breaking is a compressed spectrum of all the 4D states who descend from bulk fields~\cite{Chen:1996ap, Murayama:2012jh}.  All of the bulk states of the theory share a common $R$-symmetry; absent additional global symmetries which may be broken by boundary conditions, common boundary conditions for all of the superpartners are enforced.  Thus, up to radiative and/or volume-suppressed corrections, there will be a common mass for all zero-mode superpartners.  Of course, brane-localized states do not share this common mass, and so ultimately the implications of compression in Scherk-Schwarz theories depends on the detailed geography of brane and bulk states.
As we will discuss in Section~\ref{sec:BulkModel}, there are a variety of realistic models in which compression leads to a considerable reduction in the strength of collider limits.

\subsection{Calculability and its Discontents}

The novel properties of Scherk-Schwarz SUSY breaking have been explored in previous papers in the context of the Minimal Supersymmetric Standard Model (MSSM)~\cite{Barbieri:2000vh, Barbieri:2001dm, Barbieri:2001cz, ArkaniHamed:2001mi, Delgado:2001si, Delgado:2001ex, Barbieri:2001yz, Marti:2001iw, Delgado:2001xr, Barbieri:2002uk, Barbieri:2003kn, vonGersdorff:2003qf, vonGersdorff:2004eq, vonGersdorff:2004cg, Diego:2006py, Murayama:2012jh, Chacko:2008cb}.  In addition to a natural, calculable breaking of electroweak symmetry, these 5D implementations of the MSSM predict a relatively large Higgs mass (quite near the measured value) when one or both of the top quark multiplets propagate in the bulk.  However, since the scalar potential is controlled by only one scale,
the compactification radius is uniquely predicted by the Higgs vev, and there is no parametric freedom that could allow the tuning of radiative contributions in order to raise $1/R$ while keeping the Higgs vev fixed.  As might be expected from naturalness considerations, one then finds the observed weak scale corresponds to $1/R \simeq 700~\textrm{GeV}$~\cite{Barbieri:2000vh}, and therefore an experimentally-excluded gluino mass, $m_{\widetilde{g}} \simeq 350~\textrm{GeV}$.  

Introducing the folded sector provides the necessary parametric freedom to raise the KK mass scale beyond the reach of current bounds.  In the limit of maximal twist (see Fig.~\ref{fig:SpectrumMaxTwist} below for an illustration) the one-loop top contribution to the Higgs potential vanishes due to the presence of an ``accidental supersymmetry.''  As $\alpha$ is lowered from $\alpha = 1/2$, the size of the top contribution to the Higgs soft mass is increased from zero, while the gauge contribution is decreased, allowing the value of $1/R$ corresponding to $v \simeq 174~\textrm{GeV}$ to be continuously varied. In folded theories, the weak scale can be accommodated with larger values of $1/R$ relative to those predicted by 5D realizations of the MSSM.   

Alternatively, one could relax the requirement of calculability and couple the MSSM to arbitrary SUSY breaking on one of the branes; this possibility has recently been explored in~\cite{Dimopoulos:2014aua}. In these models, there is a new scale of SUSY breaking that is not determined by the radius of compactification. In principle, electroweak symmetry breaking along with an arbitrary spectrum can be achieved, but at the expense of introducing irrelevant operators with relatively large, unknown coefficients.  For the remainder of this paper, we will neglect this possibility by focusing on calculable scenarios in which the Scherk-Schwarz twist is the only source of SUSY breaking.

There is one caveat to complete calculable control.  Any operators allowed by $\mathcal{N} = 1$ SUSY that are consistent with all the symmetries in the theory may be generated on the 4D branes at the cutoff scale $\Lambda$.  This manifests as non-trivial brane kinetic terms and other incalculable K\"{a}hler potential operators, such as $\mathcal{K} \supset \delta(y)\, Z_{ij}\, \Phi_i^\dag \,\Phi_j$.  As the theory is strongly coupled at scale $\Lambda$, these coefficients should not be assumed to be small; in particular, they may have a substantial impact on the low-energy spectrum since $\Lambda R$ is not large. 

One concern is that the flavor structure of these terms would lead to large flavor violating effects at low energies.  Since this is set by the strong dynamics above $\Lambda$, we will assume that the UV theory respects the flavor symmetries of the Standard Model up to the Yukawa couplings, \emph{i.e.}, minimal flavor violation.  We will therefore not consider flavor constraints on the model from large brane-localized operators.  Of course, there may still be sizable flavor-preserving operators.  In general, since these operators are confined to the branes, volume suppression will render the associated corrections small.\footnote{As discussed in Appendix~\ref{app:ZFactors}, there is a possible exception when brane-kinetic terms are allowed to be negative.  In this case, the existence of KK states with zero norm at tree level necessitates working to higher order in perturbation theory.  We leave a full analysis of this scenario to future work.}   For a numerical estimate of their effects on the KK spectrum and thence the Higgs potential, see Appendix~\ref{app:ZFactors}.  

Another interesting possibility is the presence of Fayet-Iliopoulos (FI) $D$-terms on the branes.  When the hypercharges of the (tree-level) massless scalar spectrum satisfy $\sum_i Y_i \neq 0$, there is a quadratically-divergent one-loop diagram leading to incalculable FI terms localized on the orbifold fixed points~\cite{Ghilencea:2001bw, Barbieri:2001cz, Barbieri:2002ic, Barbieri:2002uk}.  These FI terms have two effects: first, they lead to explicit Higgs soft masses through the $D$-term potential; second, they quasi-localize all hypercharge-carrying states to the 4D branes through their 5D gauge kinetic terms.  The latter (subdominant) effect modifies the 5D wave functions, changing the effective couplings of these states to the Higgs through their wave function overlaps.  In principle, these effects can be included as a free, incalculable parameter in the model.  In the analysis below, we will assume the quadratically divergent contributions are not present and will set them to zero.  This is consistent with the low-energy degrees of freedom for the model in which the Higgs states are localized to the brane at $y = 0$ (see Sec.~\ref{sec:BraneModel}).  For the model with the Higgs propagating in the bulk, such a quadratic divergence does arise.  However, as we will show in Sec.~\ref{sec:BulkModel}, the light Higgs doublet lies along a $D$-flat direction; consequently, the FI terms do not contribute to its soft mass.

\section{Folded Supersymmetry}
\label{sec:fSUSY}
Having reviewed the general features of SUSY breaking by boundary conditions, this section will explain how the folded approach yields a model of neutral naturalness. We will begin with a review of the original incarnation of $f$-SUSY, which is a 5D model with maximal Scherk-Schwarz twist \cite{Burdman:2006tz}.  We will provide an estimate indicating that \f SUSY at maximal twist does not break electroweak symmetry, motivating the exploration of models with non-trivial twists.  We then discuss the general features of these models, which yield a non-zero Higgs vev at one loop.  This section provides the requisite background and motivation for the two detailed models that we will consider in Secs.~\ref{sec:BraneModel} and \ref{sec:BulkModel}.   

The essential idea underlying this avatar of neutral naturalness is the pairing of opposite-spin partner states with different Standard Model quantum numbers via discrete symmetries. Since 4D $\mathcal{N} = 1$ global supersymmetry necessarily commutes with the Standard Model gauge groups, this pairing looks like a hard breaking of SUSY.  It may, however, be successfully UV completed via a 5D supersymmetric embedding with SUSY-breaking boundary conditions.  Discrete symmetries relate bulk fields charged under different gauge interactions, ensuring the equality of their couplings.  Dimensional reduction of this theory leaves behind the desired spectrum of light states, supplemented by towers of Kaluza-Klein modes.

\subsection{Folded Supersymmetry at Maximal Twist}
\label{sec:FoldedMaxTwist}
In the minimal incarnation of \f SUSY, the bulk gauge group is $SU(2)_W \times U(1)_Y\times SU(3)_c \times SU(3)_f$.  The matter content of the MSSM is doubled; these are the $f$-states, and they carry the same charges as their MSSM partners, with the exception that the colored states are charged under $SU(3)_f$ instead of $SU(3)_c$.  A $\mathbb{Z}_2^f$ symmetry is imposed that exchanges the MSSM multiplets $\Phi$ for their folded partners $\Phi_f$.  This $\mathbb{Z}_2^f$ enforces that the Yukawa couplings to the Higgs are the same for matter and folded matter and that the $SU(3)_f$ gauge coupling is equal to $\alpha_s$ (at the scale where $\mathbb{Z}_2^f$ is a good symmetry).  Of particular importance here are the \f tops and \f stops which (along with the tops and stops) will yield the dominant contributions to the Higgs potential.  

\begin{figure}[t!]
\centering
\includegraphics[width=.65 \textwidth]{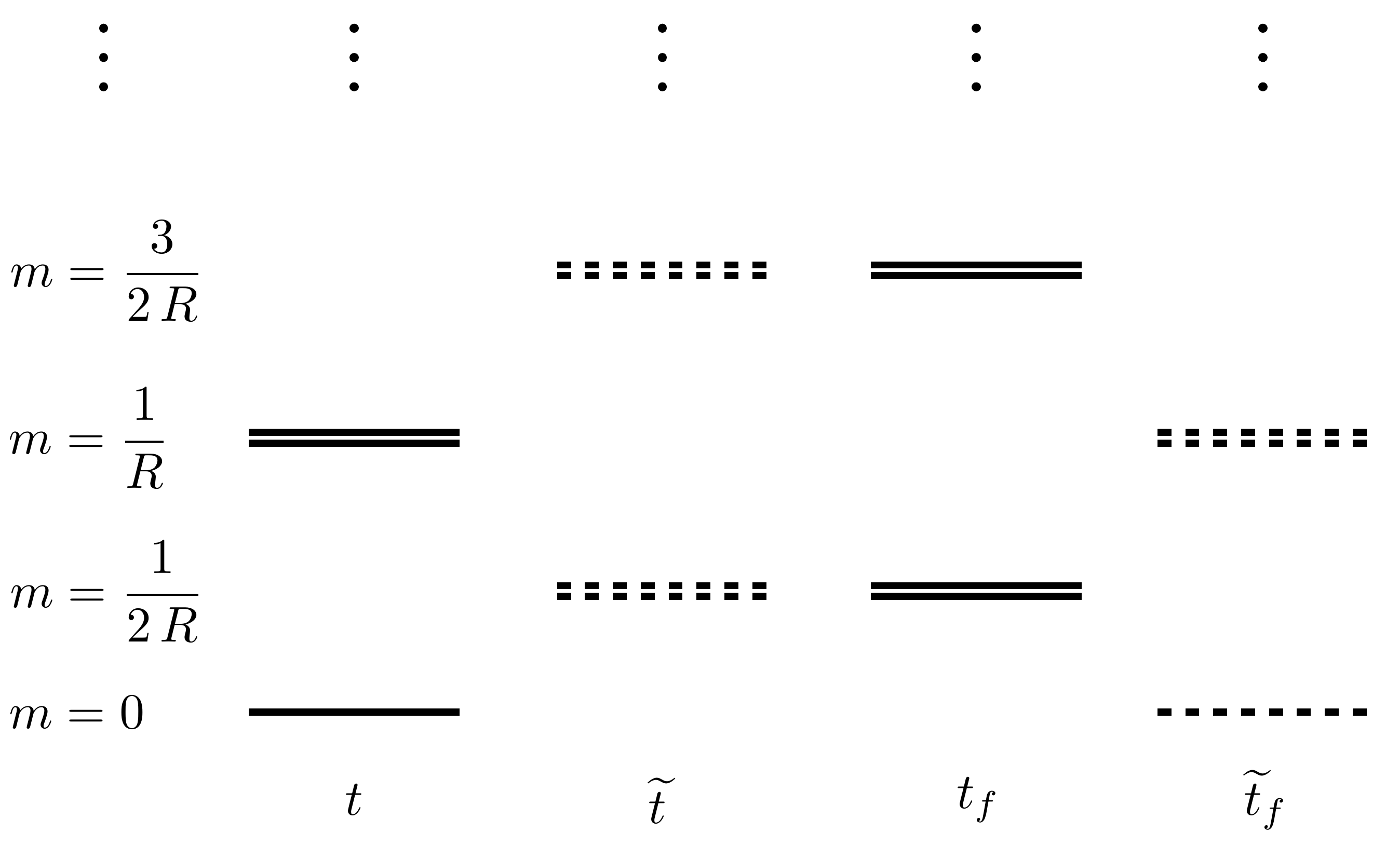}
\caption{Sketch of KK top sector spectrum at maximal twist.}
\label{fig:SpectrumMaxTwist}
\end{figure}

At maximal twist, one can label the various fields by their eigenvalues under the $\mathcal{Z}_2$ transformations that reflect $y$ about the orbifold fixed points at $y=0$ and $\pi\,R$, as defined in \eref{eq:ZandZPrime} above.  The transformation properties are given in Table \ref{tab:MaxTwistZ2s}.  The ``folded" multiplets have an inverted hierarchy between the fermionic states and their superpartners, enforced by choosing opposite eigenvalues of $\mathcal{Z}_2'$ for matter versus \f matter.  The resulting spectrum is sketched in Fig.~\ref{fig:SpectrumMaxTwist}.  Note that the gauge sector is not folded; the zero-mode gauge bosons are massless and the lightest gauginos have mass $1/(2\,R)$.

\begin{table}
\renewcommand{\arraystretch}{1.5}
\centering
\setlength{\tabcolsep}{1.5em}
\begin{tabular}{c|cccc}
matter & $\Psi\,\bm{[}+,+\bm{]}$ & $\Psi^c\,\bm{[}-,-\bm{]}$ & $\widetilde{\Psi}\,\bm{[}+,-\bm{]}$ & $\widetilde{\Psi}^c\,\bm{[}-,+\bm{]}$ \\
\hline
folded & $\Psi_f\,\bm{[}+,-\bm{]}$ & $\Psi^c_f\,\bm{[}-,+\bm{]}$ & $\widetilde{\Psi}_f\,\bm{[}+,+\bm{]}$ & $\widetilde{\Psi}_f^c\,\bm{[}-,-\bm{]}$ \\
\hline
gauge & $A^{\mu}\,\bm{[}+,+\bm{]}$ & $\Sigma\,\bm{[}-,-\bm{]}$ & $\lambda_1\,\bm{[}+,-\bm{]}$ & $\lambda_2\,\bm{[}-,+\bm{]}$ \\
\end{tabular}
\caption{Transformation properties at maximal twist.  Note $\Psi_{(f)}$ is a generic matter (\f)fermion, $\widetilde{\Psi}_{(f)}$ is a generic matter field (\f)scalar, $A^{\mu}$ is a generic gauge boson, $\Sigma$ is the associated scalar that fills out the $\mathcal{N}=2$ vector multiplet, and $\lambda_{1,2}$ are the two gauginos.  The brackets denote the choice of boundary condition for each field under $\mathcal{Z}_2$ and $\mathcal{Z}'_2$ respectively.}
\label{tab:MaxTwistZ2s}
\end{table}

The following Yukawa couplings are allowed:
\begin{align}
W \supset \delta(y)\,y_t^{(5)}\,\big(H\,Q\,U + H\,Q_f\,U_f\big),
\label{eq:TopYukawa}
\end{align}
where $y_t^{(5)}$ is the 5D top Yukawa coupling, and $\mathbb{Z}_2^f$ imposes that the Yukawa couplings are equal.  In general, Scherk-Schwarz SUSY breaking leads to tree-level non-supersymmetric splittings within bulk multiplets, which would then feed down into one-loop soft masses for brane-localized states. At maximal twist, however, the one-loop Higgs potential enjoys an apparent accidental SUSY due to ``bi-fold" protection~\cite{Burdman:2006tz} -- the scalars from the folded sector $f$-top supermultiplets are degenerate with the fermions in the Standard Model top supermultiplets (and visa versa).  This leads to a cancellation of the one-loop threshold corrections from \eref{eq:TopYukawa} for the Higgs mass at each level of the KK tower.  However, this bi-fold protection holds only for the contribution of top-stop loops to the Higgs doublet masses. Gauge invariance of the Yukawa couplings requires all states to be charged under a single $SU(2)_W \times U(1)_Y$, and so there can be no doubling in the gauge sector.  Accordingly, electroweak gauginos contribute to the Higgs mass at one loop, as in the 5D MSSM.  Thus at maximal twist the Higgs potential accumulates one-loop contributions from the gauge-gaugino sector but not the stop-top sector.

The squarks likewise receive soft masses at one loop from their gauge and Yukawa couplings. In particular, the (\f)stop soft masses are~\cite{Burdman:2006tz}
\begin{align}
\delta m_{\widetilde{Q}}^2 &= K\frac{1}{4\, \pi^4}\frac{1}{R^2}\left(\frac{1}{2}\,y_t^2 + \frac{4}{3}\, g_3^2 + \frac{3}{4}\, g_2^2+\frac{1}{36}\, g_1^2\right) \simeq 0.015 \frac{1}{R^2},\nonumber\\
\delta m_{\widetilde{U}}^2 &= K\frac{1}{4\, \pi^4}\frac{1}{R^2}\left(y_t^2 + \frac{4}{3}\, g_3^2 +\frac{4}{9} \,g_1^2\right)  \simeq 0.016 \frac{1}{R^2},
\label{eq:1LoopStopMass}
\end{align}
where $K = 7\, \zeta(3)/4 \simeq 2.1$ and an exact $\mathbb{Z}_2^f$ symmetry on the values of the couplings has been assumed for the numerical evaluation. 

At two loops, these SUSY-breaking masses feed into the Higgs potential.   It is reasonable to assume that the dominant effect of this two-loop soft-breaking of SUSY can be captured by including the one-loop soft masses as a shift in the pole of the propagator for the relevant states.\footnote{This is in close analogy with ``daisy resummation" in finite temperature field theory (see \emph{e.g.} \cite{Quiros:1999jp} for a nice review), except here we are compactifying a spatial dimension from 5D to 4D as opposed to compactifying the time direction from 4D to 3D.}   Note that we model this effect as a universal shift in the mass of the entire tower of (\f)top squarks, which is a reasonable approximation given the explicit values shown in Eqs.~(\ref{eq:1LoopStopMass}).  Then the contribution to the Higgs mass squared parameter can be computed using standard 5D techniques, as described in Appendix~\ref{app:Potential}.  The result is
\begin{align}
m_H^2 = \mu^2 + \frac{21\, \zeta (3) \left(g^2+g'^2/3\right)}{64 \, \pi ^4\, R^2}+\lambda _t^2\, m_{\tilde{t}}^2 \left(\frac{3  \left(\log \big[(4\, \pi\, R\, m_{\tilde{t}})^2  \big]-3\right)}{8\, \pi ^2} + \cdots \right),
\label{eq:HiggsMassMaximalTwist}
\end{align}
where $\mu$ is the standard MSSM Higgsino mass, the second term is due to the one-loop contribution from gauge/gaugino loops, and the third term comes from our approximate evaluation of the leading two-loop effects.  We denote by $m_{\tilde{t}}$ the universal one-loop stop/$f$-stop soft mass in the approximation that all the stops and $f$-stops receive identical radiative corrections to their masses.  

Using \eref{eq:HiggsMassMaximalTwist}, it is straightforward to see that achieving $m_H^2 < 0$ requires $m_{\tilde{t}}^2 \gtrsim 0.04/R^2$ when
 $\mu = 0$ (which would be excluded by chargino limits if the Higgs is brane-localized, but is allowed if the Higgs propagates in the bulk and the Higgsino is rendered massive by Scherk-Schwarz boundary conditions);  $m_{\tilde{t}}^2$ needs to be strictly larger than this for $\mu \neq 0$.  Combined with Eqs.~(\ref{eq:1LoopStopMass}), this provides evidence that the model as originally constructed does not break electroweak symmetry.  Specifically, the one-loop stop/$f$-stop mass would need to be a factor of $\sim3$ larger before this two-loop effect would overwhelm the one-loop soft mass from the gauge sector and drive electroweak symmetry breaking.

We note that our approximate two-loop contribution in \eref{eq:HiggsMassMaximalTwist} differs from the two-loop estimate in \cite{Burdman:2006tz}, which was used to argue for the viability of electroweak symmetry breaking at maximal twist. The estimate in \cite{Burdman:2006tz} was performed by running the Higgs mass from the (\f)top sector states starting at the scale $1/R$, yielding a dependence on $\log(R\, m_{\tilde t})$.  However, the calculation involving the full sum over KK modes depends on $\log(4\, \pi\, R \, m_{\tilde t})$, which proves critical for overcoming the positive gauge contribution.   This can be understood from the five-dimensional perspective -- the argument of the logarithm should be a function of $2 \, \pi\, R$, rather than $R$, because loops must propagate a distance $2\, \pi\, R$ from one boundary to the other and back again in order to see that SUSY is broken.  Note that the finite terms in \eref{eq:HiggsMassMaximalTwist} also play a numerically non-trivial role.
 
There are a couple of important caveats to this claim.   It is possible that additional corrections from the full two-loop calculation could yield the proper vacuum.  Furthermore, this calculation assumes that the incalculable, brane-localized K\"{a}hler potentials for the top and \f top superfields are negligible.  Nevertheless, we take this as motivation to extend the model in order to achieve electroweak symmetry breaking at one loop.

\subsection{Folded Supersymmetry at Arbitrary Twist}
The choice of $\alpha = 1/2$ made the top quark and $f$-stop zero modes degenerate at tree level, thereby endowing $f$-SUSY with bi-fold protection (see Fig.~\ref{fig:SpectrumMaxTwist}).  By moving $\alpha$ away from $1/2$, the degeneracy is broken, and with it, the bi-fold protection of the Higgs potential.

The twist can be formulated as follows.  There are two possible actions for 5D orbifolds: reflection $\mathcal{Z}$ about $y = 0$, and translation $\mathcal{T}$ by $2\,\pi\,R$.  These can be identified with the $\mathcal{Z}$ reflection about $y=0$ and the $\mathcal{Z}' = \mathcal{T}\,\mathcal{Z}$ reflection about $y= \pi\,R$ defined in \eref{eq:ZandZPrime}. Discussing the twist in the language of the translation is more natural from the perspective of action on superfields which are not eigenstates of $\mathcal{Z}_2'$.

For the matter fields, the $\mathcal{Z}_2'$ acts as
\begin{align}
\mathcal{Z}_2'
\left( \begin{array}{c}
\Phi \\
\Phi^{c\dag}
\end{array}
\right) (y)
= \left( \begin{array}{c}
\Phi \\
\Phi^{c\dag}
\end{array}
\right) (2\,\pi\,R-y)
= \mathcal{T}\,\cdot\, \sigma_3 \,\cdot
\left( \begin{array}{c}
\Phi \\
\Phi^{c\dag}
\end{array}
\right)(y),
\label{eq:nonMaxZ2prime}
\end{align}
where 
\begin{align}
\mathcal{T} = \left \{
\begin{array}{cc}
e^{-2\,i\,\pi\,\alpha \,\sigma_2} (-1)^f \quad\quad & \text{for } SU(2)_R \text{ doublet}\\
\mathbb{1}  & \text{for } SU(2)_R \text{ singlets}
\end{array}
\right. , 
\label{eq:TwistedBC}
\end{align}
and $f = 0\,(1)$ for matter (folded) fields.  The gauginos transform as do the matter fields in \eref{eq:TwistedBC}, while the $A^\mu$ and $\Sigma$ transform as in Table~\ref{tab:MaxTwistZ2s}.  The transformation properties of bulk Higgses are given in Appendix~\ref{app:wavefunctions}.

\begin{figure}[t]
\centering
\includegraphics[width=.65 \textwidth]{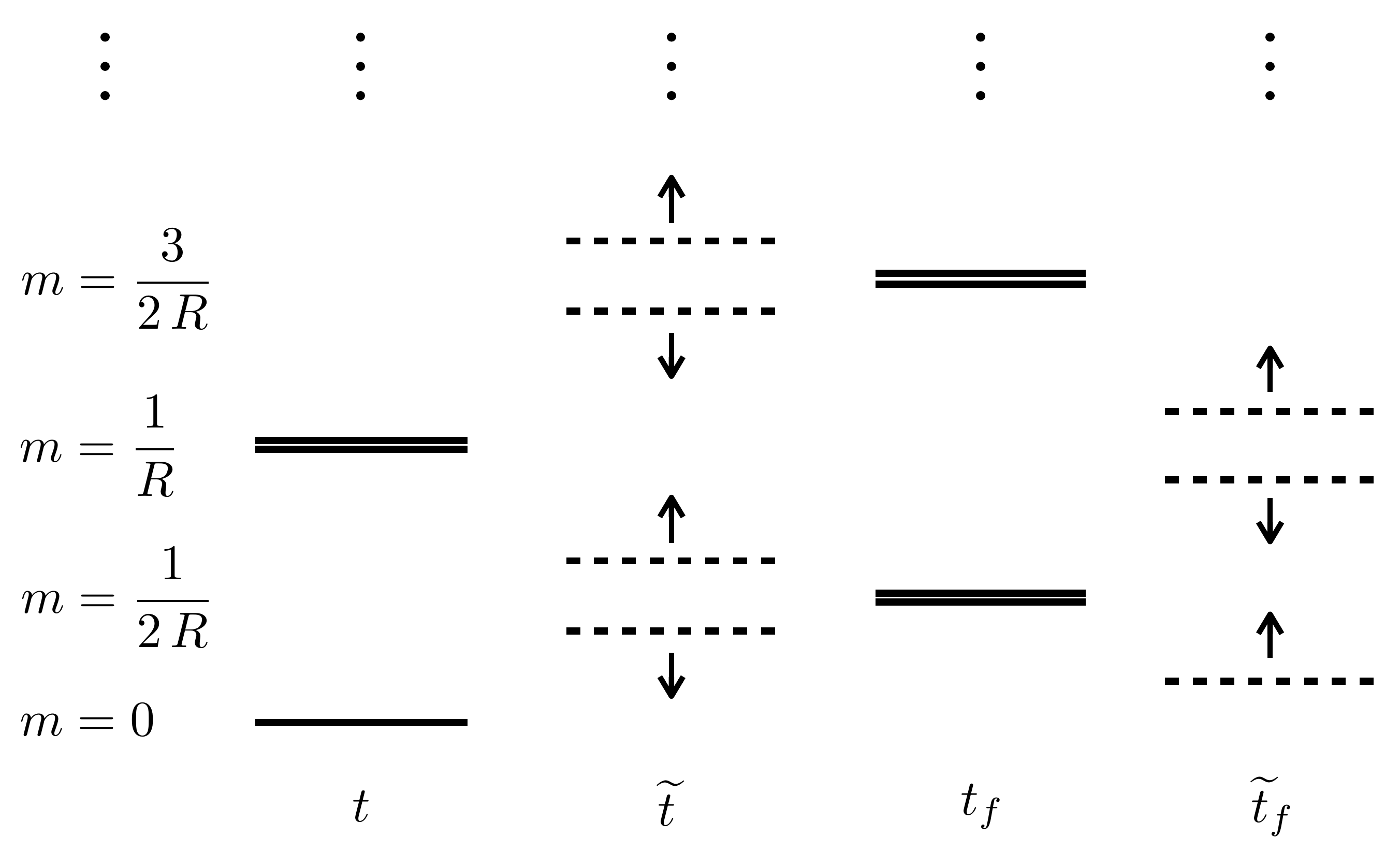}
\caption{Sketch of KK top sector spectrum at arbitrary twist $\alpha$.  The arrows indicate how the masses of the states change as $\alpha$ deviates from $1/2$.}
\label{fig:SpectrumArbitraryTwist}
\end{figure}

After electroweak symmetry breaking, the combination of the non-maximal twist and the top Yukawa on the $y=0$ brane yields kinked KK wave functions for the stops:
\begin{align}
\left( \begin{array}{c}
\widetilde{t} \\
\widetilde{u}^c \\
\widetilde{u}^\dag \\
\widetilde{t}^{c \dag}
\end{array} \right)_{\pm}
= 
\frac{1}{\sqrt{4\,\pi\,R}} \sum_n
\left( \begin{array}{c}
\cos \left(\frac{k_n \, \{y\} + \alpha \,y}{R} - \pi k_n\right) \\
\pm \sin \left(\frac{k_n \, \{y\} + \alpha\, y}{R} - \pi k_n\right) \\
\pm \cos \left(\frac{k_n \, \{y\} + \alpha \,y}{R} - \pi k_n\right) \\
\sin \left(\frac{k_n \, \{y\} + \alpha \, y}{R} - \pi k_n\right)
\end{array} \right)_{\pm}
\widetilde{t}^{\,(n)}_{\pm},
\label{eq:stopWF}
\end{align}
with $\{y\} \equiv y \mod 2\, \pi\, R$, such that $\{y\}$ is restricted to the fundamental domain.  Here $k_{n\pm}~\equiv~n~+~(1/\pi)\tan^{-1}(\pm\,\pi \, R \, y_t \, v)$,  $n \in \mathbb{Z}$, $y_t$ is the 4D top Yukawa coupling, and $v \simeq 174 \GeV$ is the Higgs vev.  The wave functions for the $f$-stops are given by taking \eref{eq:stopWF} and replacing $\alpha \rightarrow \alpha -1/2$.  These wave functions can then be used to compute the Coleman-Weinberg potential from top/stop and $f$-top/$f$-stop loops.  The resulting tree-level masses for the KK towers of (s)tops and $f$-(s)tops are
\begin{align}
\begin{aligned}
m_t^{(n)} = \bigg|\frac{k_n}{R}\bigg|  \quad\quad&\quad\quad m_{\tilde{t}}^{(n)} = \bigg|\frac{k_n+\alpha}{R}\bigg|\\[10pt]
m_{t_f}^{(n)} = \bigg|\frac{k_n+1/2}{R}\bigg|  \quad\quad&\quad\quad m_{\tilde{t}_f}^{(n)} = \bigg|\frac{k_n+1/2-\alpha}{R}\bigg|.
\end{aligned}
\end{align}
This spectrum is sketched in Fig.~\ref{fig:SpectrumArbitraryTwist}.

Since it acts differently on $SU(2)_R$ doublets and singlets, the non-trivial twist also impacts the gaugino masses. For every vector supermultiplet, there is a zero-mode Majorana gaugino $\lambda^{(0)}$ of mass $m_\lambda^{(0)} = \alpha/R$, plus towers of Majorana gauginos $\lambda^{(n)}$ of masses $m_\lambda^{(n)} = |n + \alpha| / R$.   These towers arrange themselves into Dirac gauginos at maximal twist. The lightest gauginos consist of the usual Majorana bino, wino, and gluino, plus an $f$-gluino, all with (tree-level) mass $\alpha / R$. 

In contrast to maximal twist, these $\alpha$-dependent, SUSY-breaking mass splittings allow for electroweak symmetry breaking at one loop.  However, this comes at a price. While \f SUSY at maximal twist had the virtue of lifting all new colored states to $\sim \frac{1}{2R}$ and preserving naturalness through light $f$-stops, \f SUSY at non-maximal twist features colored states at the scale $\sim \frac{\alpha}{R}$ with $\alpha < 1/2$. For a given radius of compactification, this brings colored states closer to the weak scale, where strong LHC limits enter to provide relevant constraints. As we will see in the following sections, it will not be possible to achieve the measured Higgs vev and mass without falling afoul of existing phenomenological constraints, and so some additional ingredients are required.  Before proceeding to the complete models, we will first make a few generic comments about \f SUSY phenomenology.

\subsection{Folded Phenomenology}
\label{sec:FoldedPheno}
Folded SUSY contains all of the phenomenology of the MSSM, along with a very rich set of observables from the folded sector.  Given the choice of requiring calculable SUSY breaking from extra dimensions, we will demonstrate below that the dominant constraints on the parameter space will be due to standard MSSM-like signatures:  Higgs mixing and heavy Higgs production for the brane Higgs model (see Sec.~\ref{sec:BraneModel}) and gluino-squark limits for the bulk Higgs model (see Sec.~\ref{sec:BulkModel}).  Generically, twisted \f SUSY models predict an MSSM-like discovery immediately followed by the observation of folded-sector states.  The purpose of this section is to briefly review some of these observables, along with some additional complications that are required to make \f SUSY compatible with cosmology.

Recall that the folded states couple to the Standard Model via the electroweak gauge group along with their Yukawa couplings to the Higgs (this is only relevant for the 3$^\text{rd}$ generation) -- thus folded states may be produced via either electroweak processes or the Higgs portal.  Since there are no light fermions in the folded sector, the folded phenomenology is very different from that of the visible sector.  

The phenomenology of $f$-squark production at hadron colliders has been studied in detail in \cite{Burdman:2006tz, Harnik:2008ax, Burdman:2008ek, Burdman:2014zta, Curtin:2015fna}; here we briefly summarize the relevant physics and make several new observations. When pairs of $f$-squarks are produced at the LHC, the lack of light fermions implies that the color string (with associated confinement scale $\Lambda_f \sim \Lambda_\text{QCD}$) connecting these states cannot be broken by pair production from the vacuum.  The result is that the $f$-squark pairs rapidly radiate $f$-glueballs and soft photons before eventually annihilating. The dominant annihilation products depend sensitively on the properties of the $f$-squark pair. Neutral-current production of light-flavor $f$-squarks typically ends in annihilation to twin glueballs, while charged-current production typically ends in annihilation to $W^\pm \gamma$. 

The situation is somewhat different for $f$-stops and $f$-sbottoms. If the $f$-stops are heavier than the $f$-sbottoms, they typically undergo $\beta$-decay before annihilating, and the phenomenology is similar to that of light-flavor $f$-squarks. However, if the $f$-stops are lighter, then the $f$-sbottoms undergo $\beta$-decay into the $f$-stops, and annihilation may proceed into final states involving Higgs bosons. Although the spectrum of $f$-squarks is nominally fixed in terms of $\alpha$ and $R$, the effects of incalculable brane-localized wave function renormalization factors can yield a re-shuffling of the $f$-squark spectrum, so that all possible orderings should be considered. Thus the generic signatures of $f$-squark pair production are resonances in $V\,V'$ channels where $V,\, V' = W^\pm,\,Z^0,\,\gamma,\,h, \, g_f$, depending on the identity of the $f$-squark pair.  Note that the timescale for this annihilation is prompt at the LHC.  Limits have been estimated for these channels, leading to a bound of $m_{\tilde q_f} \gtrsim 300-500$ GeV from $W^\pm \, \gamma$ resonance searches \cite{Burdman:2014zta}. Although no current bound has been placed using $W^\pm\, h$ and $h\,h$ resonance searches, these should provide interesting limits as well. Decays entirely into $f$-glueball states followed by displaced decay to light Standard Model states are also likely to prove promising \cite{Curtin:2015fna}.

There is also phenomenology associated with the $f$-sleptons.  In particular, the lightest $f$-slepton is stabilized by $f$-lepton number and, in some cases, $R$-parity conservation.  If it is electromagnetically charged, there are strong cosmological constraints.  A simple way to avoid such constraints is to add additional terms to the superpotential that violate these symmetries.  For example, including $y^d_{i j}\, L_f\,Q_i\,D_j$ in the superpotential on the $y=0$ brane would allow the $f$-sleptons to decay to di-jets.  The Yukawa coupling is included so as to avoid spoiling the MFV assumption for the MSSM states.  As the \f slepton must only decay before BBN, the residual flavor violation from the \f lepton sector can be made negligible.  Despite the $R$-parity violation, the lightest neutralino remains stable, as all lighter fermions carry either Standard Model baryon or lepton number, which are unbroken by the addition of this operator.

Alternatively, another version of \f SUSY could be constructed where the $\mathbb{Z}_2^f$ symmetry only applies to colored states, such that there is no doubling of bulk lepton supermultiplets. In this case, the zero-mode spectrum remains anomaly-free despite the elimination of folded lepton supermultiplets as all $f$-leptons are made massive from boundary conditions even when the lepton sector is folded. The only possible consequence of the missing matter is the introduction of bulk Wess-Zumino consistency terms \cite{Barbieri:2002ic}, which have no significant impact on the phenomenology discussed here. 

\section{Higgs on the Brane}
\label{sec:BraneModel}
\begin{figure}[t!]
\centering
\includegraphics[width=.6 \textwidth]{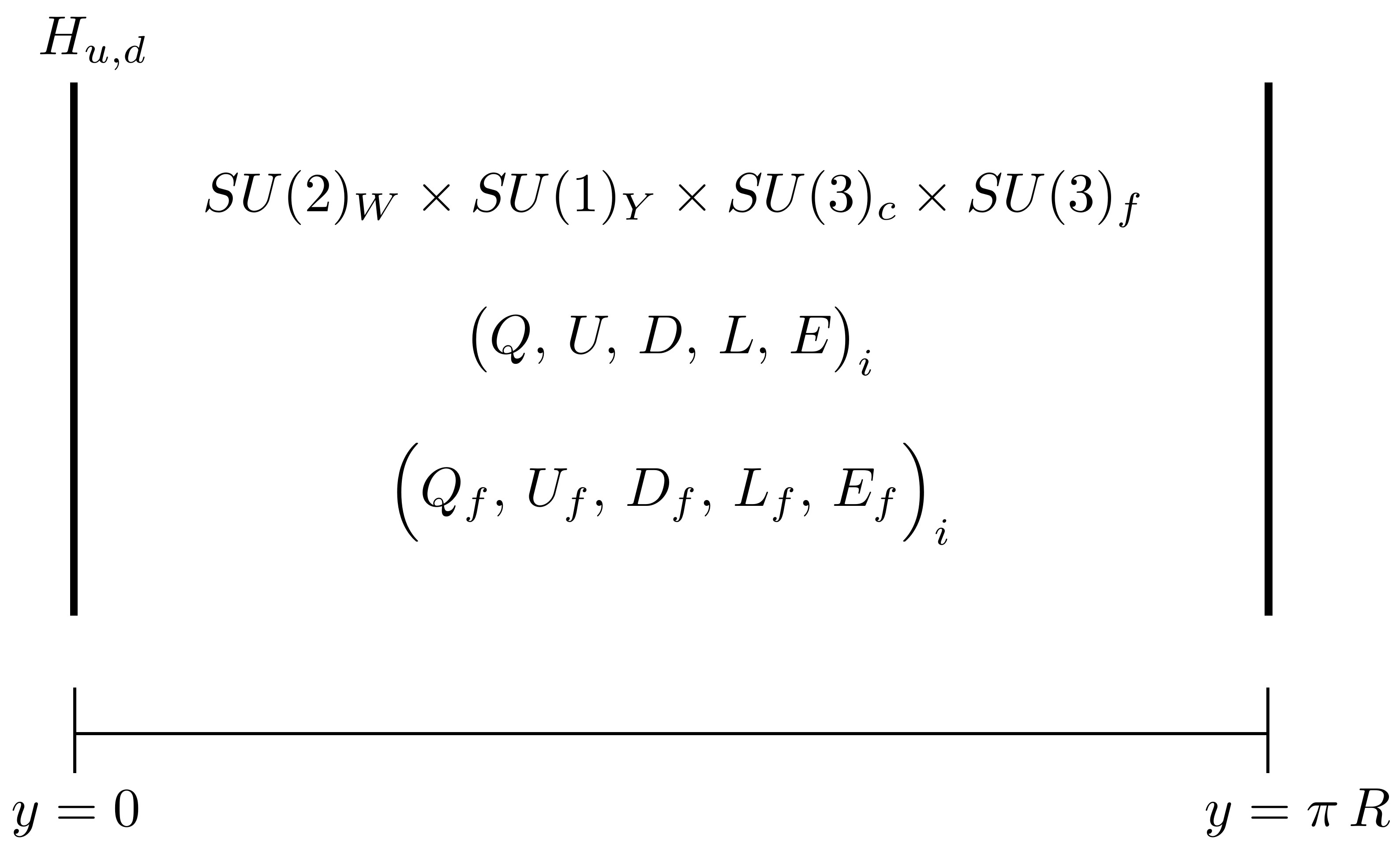}
\caption{A sketch of the geography for the brane Higgs model.  The standard MSSM two Higgs doublets $H_{u,d}$ live on the $y=0$ brane, while the gauge groups, matter, and $f$-matter fields are in the bulk.}
\label{fig:Brane_Geo}
\end{figure}

Here we consider the field configuration of the original model of \f SUSY: the matter and gauge fields are taken to propagate in the bulk, so that their zero-mode superpartners are all lifted by Scherk-Schwarz SUSY breaking, while the Higgs doublet chiral superfields $H_{u,d}$ are confined to the $y=0$ brane.  This setup is sketched in Fig.~\ref{fig:Brane_Geo}.  As in the MSSM, a $\mu$-term is required to lift the Higgsinos above current bounds, leading to an irreducible, tree-level tension between a large Higgsino mass and a weak-scale Higgs vev.  This is in addition to the challenges of electroweak symmetry breaking which, as discussed in Sec.~\ref{sec:FoldedMaxTwist} above, motivate extending the model to arbitrary twist parameter $\alpha$.  

However, in the absence of additional ingredients, the regions of parameter space that break electroweak symmetry are excluded by constraints on the Higgs sector.  Higgs coupling predictions run afoul of precision measurements at the LHC, and the extra MSSM Higgs states are excluded by direct searches.  The heart of the issue is that the leading contribution to the $b_\mu$ term, given in Eq.~(\ref{eq:VBraneHiggsExpanded}), arises at one loop via the combination of the brane $\mu$ term and the $\mathcal{O}(\alpha/R)$ effective $A$-term induced by the Scherk-Schwarz twist. Since $\mu$ must itself be $\mathcal{O}(\frac{y_t}{4 \pi} \frac{\alpha}{R})$ for viable electroweak symmetry breaking, this implies $b_\mu$ is suppressed by an amount $\sim y_t / 4 \pi$ relative to other masses-squared in the Higgs sector. Similarly, the one-loop contribution to the down-type Higgs mass-squared is proportional to the electroweak gauge couplings and suppressed relative to the up-type Higgs mass-squared by an amount $\sim y_t^2 / g^2$.

One minimal solution for lifting the second Higgs doublet without introducing new sources of SUSY breaking is to introduce the leading ``BMSSM" operator which appears at dimension-5 in the superpotential \cite{ArkaniHamed:2001mi, Dine:2007xi}. Supersymmetry is still broken strictly by boundary conditions; the new contribution acts as an effective $b_\mu$ term after electroweak symmetry breaking. The BMSSM operator also modifies the quartic of the light Higgs, but in most of the viable parameter space this is a small correction -- the dominant correction to the tree-level prediction for the Higgs mass comes from the radiative quartic coupling.  

The complete viable model is given by the brane superpotential
\begin{align} \label{eq:WBraneHiggs}
\mathcal{W} = \delta(y)\left(\mu\,H_u\,H_d - \frac{1}{M}\left(H_u\,H_d\right)^2 + y_t^{(5)}\,H_u\,(Q\,U+Q_f\,U_f) \right),
\end{align}
from which we calculate the Higgs potential in Eq.~(\ref{eq:VBraneHiggsExpanded}), including the leading radiative corrections.  The full one-loop Coleman-Weinberg contribution from the (\f)top sector has been included, while the one-loop contributions to the Higgs soft mass from the $SU(2)_W\times U(1)_Y$ gauge bosons and the leading $b_\mu$-term are computed at fixed-order.  These varying levels of approximation can be justified as follows:  the presence of the Higgs vev on the $y=0$ brane breaks both the 5D Lorentz invariance as well as the brane interchange parity.  In the 4D theory, this leads to non-conservation of both KK number and KK parity, with the result that the number of sums over KK modes grows with the number of external Higgs legs.  Therefore, it is important to include the full resummed Coleman-Weinberg potential for any contributions that are not already present at tree level.  There is a tree-level contribution to the quartic from gauge couplings, and to $H_u-H_d$ mixing from the BMSSM terms; therefore, we only include the fixed-order contributions from these sources.  The resulting potential is
\begin{align}
\label{eq:VBraneHiggsExpanded}
V =&\,  -\frac{9\, \Re\left[
\Li_5(e^{4\,i \,\pi (\alpha+\gamma)}) + \Li_5(e^{4\,i \,\pi (\alpha-\gamma)}) - 2\Li_5(e^{4\,i \,\pi \,\gamma})
\right]
}{512\, \pi^6 \,R^4 }\notag\\[10pt]
&+ \left(\mu^2 +\frac{3\,g^2 + g^{\prime 2}}{16\, \pi^4\, R^2} \Re\left[\zeta(3) - 
\Li_3(e^{2\,i\,\pi\, \alpha})\right] \right)\, \Big(|v_u|^2 + |v_d|^2\Big)  \notag \\[10pt]
 & -\frac{3\, y_t^2\, \mu\, \Im[\Li_2(e^{4\,i\,\pi\, \alpha})]}{16\, \pi^3\, R} \big(v_u \,v_d + \text{h.c.}\big) \notag \\[10pt]
 & + \frac{g^2+g'^2}{8}(|v_u^2|-|v_d|^2)^2 - 2 \frac{\mu}{M}\Big(|v_u|^2+|v_d|^2\Big)\big(v_u\,v_d + \text{h.c.}\big),
\end{align}
with $\vev{H_{u,d}} = v_{u,d}$ and $\gamma = \frac{1}{\pi} \tan^{-1}(\pi\, R \,y_t\, v_u)$.  Note that in the plots, we have also included the $\mathcal{O}\left(1/M^2\right)$ contributions to the potential, which are omitted here for brevity.  The effects of these terms on the numerical results are minor.

\begin{figure}[t!]
\centering
\quad
\includegraphics[trim=15mm 10mm 20mm 20mm, width=.45 \textwidth]{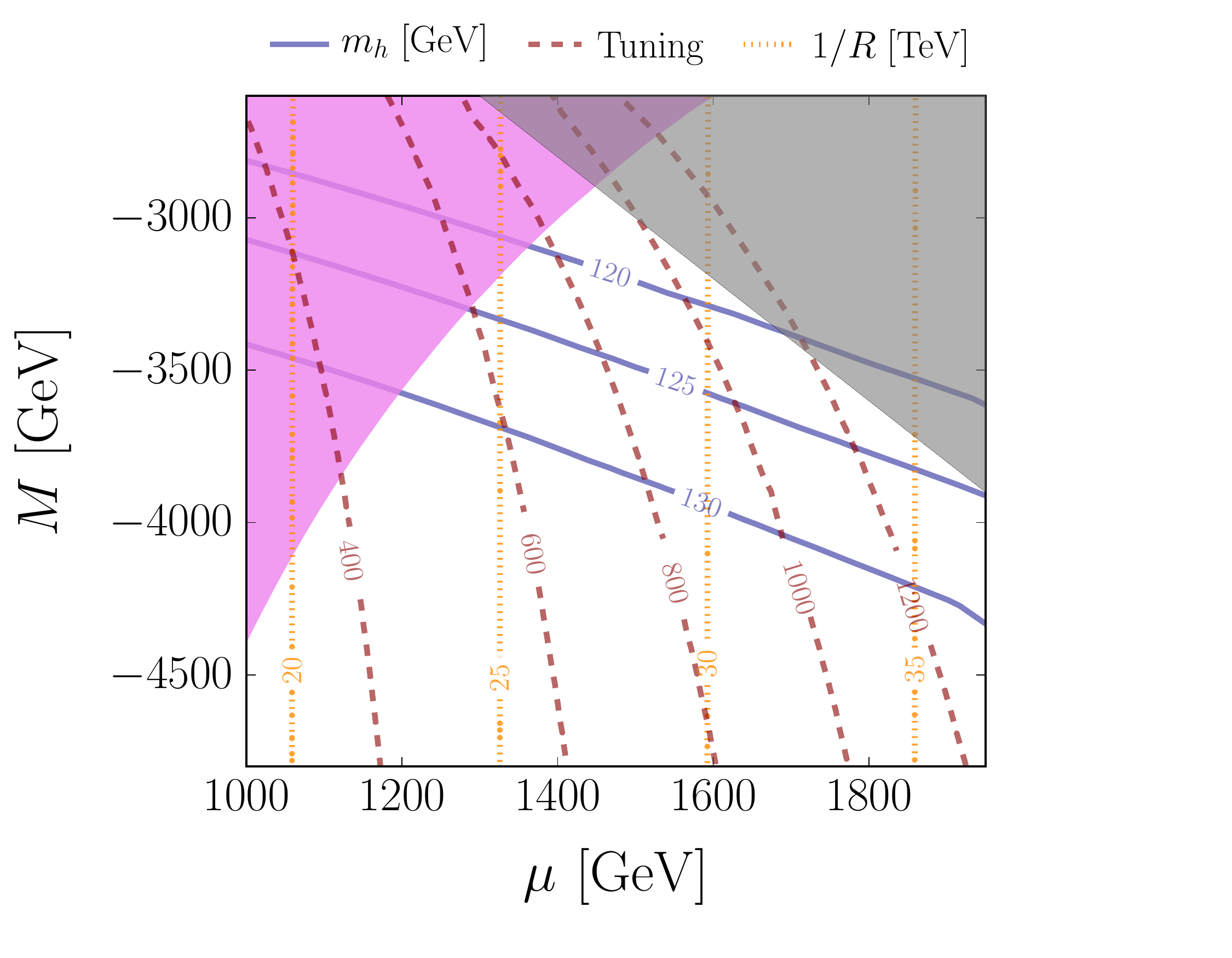}
\qquad
\includegraphics[trim=15mm 10mm 20mm 20mm, width=.45 \textwidth]{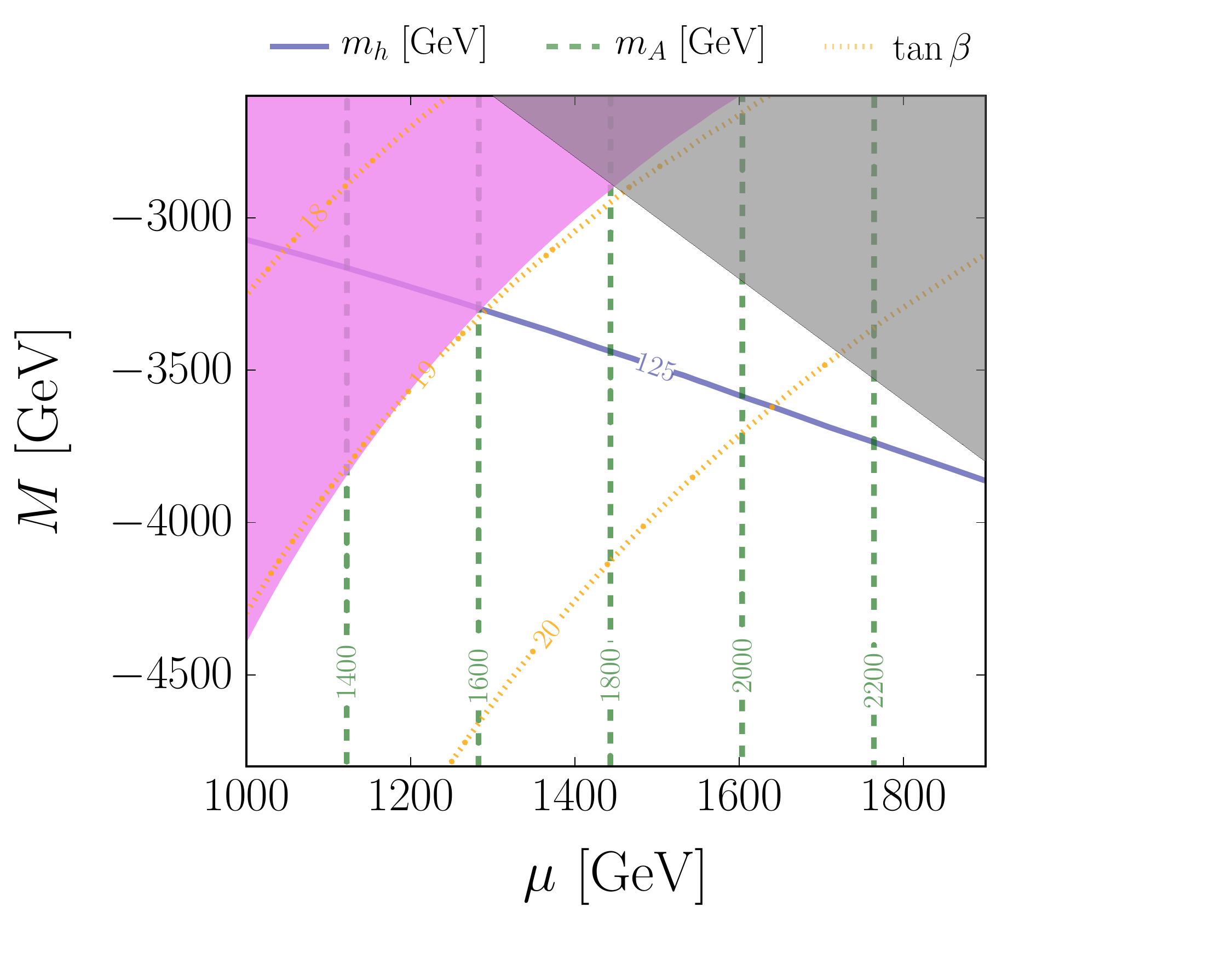}
\caption{A slice of the brane Higgs model parameter space in the $M$ versus $\mu$ plane for $\alpha = 1/3$. The left panel shows contours of fine-tuning [dashed red], $1/R$ [dotted orange], and the SM-like Higgs boson mass [solid blue]. The right panel shows contours of $\tan \beta$ [dotted orange], and the pseudo-scalar Higgs mass $m_A$ [dashed green].  The purple region is excluded by Higgs coupling measurements and the gray shaded region denotes $2\mu > |M|$, where the BMSSM approximation begins to break down.  The minimum allowed tuning compatible with a 125 GeV Higgs mass is $\sim0.2\%$.}
\label{fig:BraneHiggsParamSpace_mATanBeta}
\end{figure}

Electroweak symmetry breaking compatible with existing constraints can now be realized. In Fig.~\ref{fig:BraneHiggsParamSpace_mATanBeta} we present a slice of the parameter space in the $M$ versus $\mu$ plane, fixing $\alpha = 1/3$.  The left panel shows contours of $1/R$, fine-tuning,\footnote{Fine-tuning is computed using the standard Barbieri-Giudice measure $\Delta \equiv \text{max}_i(\text{d} \log v^2/\text{d} \log X_i)$, where $X = \{\mu, y_t, g, \alpha, M \}$~\cite{Barbieri:1987fn}.} and the physical Higgs boson mass, while the right panel shows contours of the pseudo-scalar Higgs mass $m_A$, and $\tan \beta \equiv v_u/v_d$.  The region excluded by constraints from Higgs coupling measurements is shaded purple. The region with $2 \mu > |M|$ is shaded in grey, in order to demarcate the region where the BMSSM approximation is expected to break down. Given that the BMSSM term only serves as an effective parameterization of possible extensions of the Higgs sector, there should in principle be additional light states that contribute to the Higgs potential in this gray region. Since the tuning is  larger here, we do not pursue the details of EWSB for these parameters.

The dominant limits in this model are due to Higgs mixing constraints, with all parameter space in the purple region excluded by Higgs coupling measurements.  These constraints exclude even larger values of $m_A$ than they would in the MSSM.  Although the BMSSM operator in \eref{eq:WBraneHiggs} allows the pseudoscalar mass $m_A$ to be raised, it does so in a way that slows the approach to the decoupling limit relative to the MSSM. Increasingly stringent limits on Higgs coupling deviations, largely driven by ATLAS Higgs coupling measurements, translate into a substantial bound on the scale of additional Higgs states in this model. To characterize these bounds, we perform a combined fit to ATLAS, CMS, D$\emptyset$, and CDF Higgs measurements using the package \texttt{HiggsSignals v.1.3.2}~\cite{Bechtle:2013xfa}, parameterizing deviations in terms of the tree-level mixing angles in a Type 2 2HDM.  Here we include the effects on loop-level couplings from tree-level mixing but no additional loop contributions from new states. We translate these bounds into a 95\% exclusion in the $M$ versus $\mu$ plane, where the tree-level mixing angles are computed numerically at each point in parameter space. 

In principle, the parameter space is also constrained by direct search limits on heavy scalar and pseudoscalar Higgs states, which extend to $\sim 1$ TeV at large $\tan \beta$ \cite{Aad:2014vgg, Khachatryan:2014wca}. However, the viable parameter space for EWSB falls at moderate values of $\tan \beta$ with $m_A \gtrsim 1.5$ TeV already. Constraints on MSSM superpartners could also be relelvant, but given the large $1/R \gtrsim 25$ TeV, these constraints are subdominant.
The result is that \f SUSY with brane-localized Higgses is tuned at around the  0.2\% level. In the least-tuned regions of parameter space compatible with Higgs coupling measurements, the $f$-stop masses are on the order of 8 TeV. As such, the first signatures would arise in the form of Higgs coupling deviations. The heavy MSSM-like states and the more exotic folded-sector phenomenology will be of order $8$ TeV and $4$ TeV, and will be out of reach of 13 TeV LHC searches.

While the parameter space of the brane Higgs model is strongly constrained by Higgs couplings, \f SUSY with a twist is sufficient to put superpartners out of reach of current LHC searches.  Of course, additional sources of brane-localized SUSY breaking would allow the new states in the Higgs sector to be decoupled without substantially increasing electroweak fine-tuning, albeit at the cost of reduced predictivity.

\section{Higgs in the Bulk}
\label{sec:BulkModel}
\begin{figure}[t!]
\centering
\includegraphics[width=.6 \textwidth]{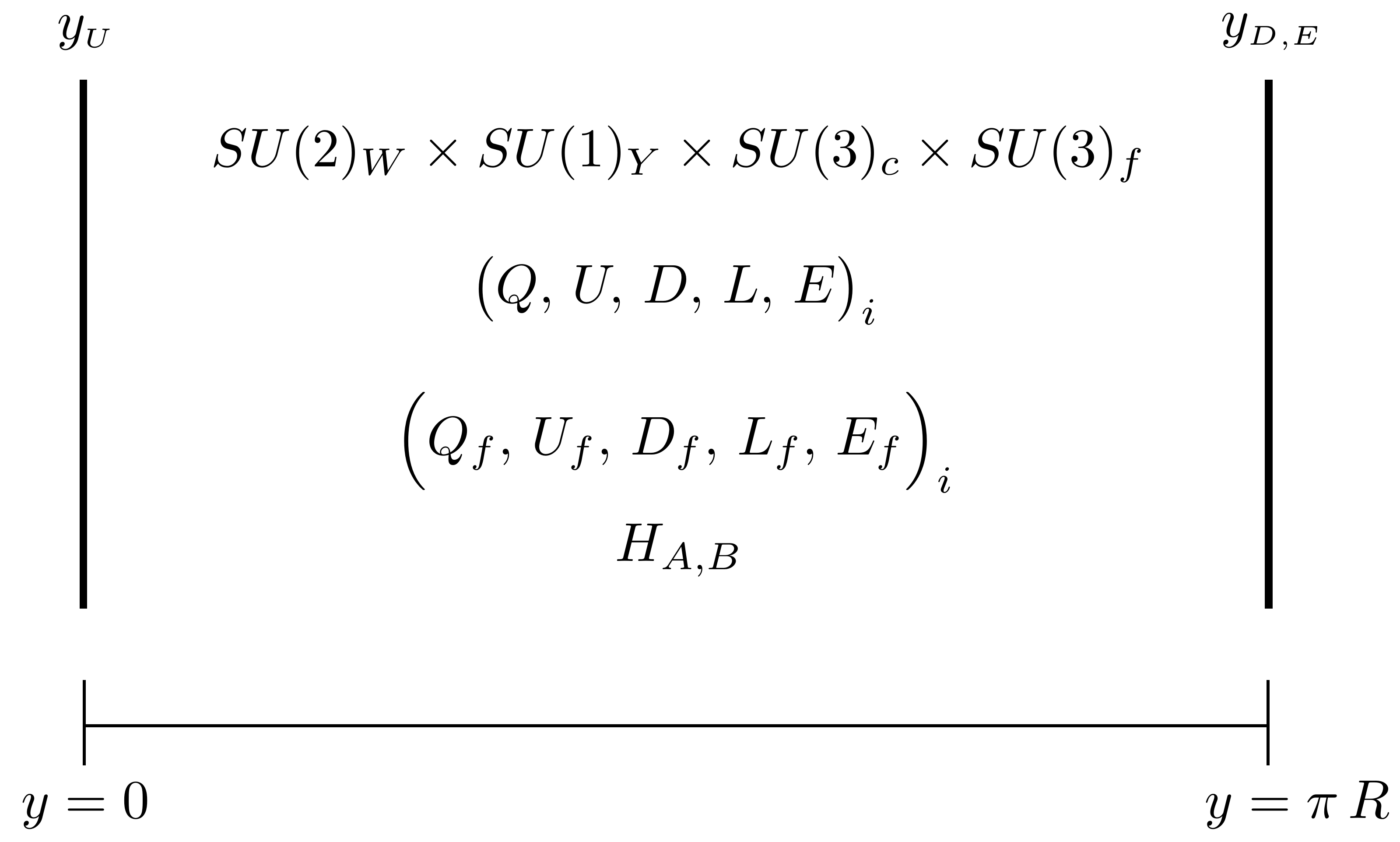}
\caption{A sketch of the geography for the bulk Higgs model.  The Higgs doublets $H_{A,B}$, gauge groups, matter, and $f$-matter fields are all bulk fields.  The up-type (down-type) Yukawa couplings are on the $y=0$ ($y=\pi\,R$) brane.}
\label{fig:Bulk_Geo}
\end{figure}

The experimental tensions discussed in the previous section motivate exploring models where the Higgs propagates in the bulk, since the choice of boundary conditions that yields a light scalar Higgs is compatible with one that lifts the Higgsinos and other Higgs scalars to scales $\sim 1/R$.  As we will discuss below, a pair of bulk hypermultiplets $H_{A,B}$ will be required.  A sketch of the geography for this model is given in Fig.~\ref{fig:Bulk_Geo}.  Note that in this setup, the up-type Yukawa is confined to the $y=0$ brane, while the down-type Yukawa couplings are located on the $\pi\,R$ brane $\big($and involve $H_B^c$, \emph{e.g.} $\mathcal{W} \supset \delta(y)\,y_b^{(5)}\,H^c_B\,Q\,D$$\big)$.

Following \cite{Pomarol:1998sd, Barbieri:2000vh}, a natural choice would be to attempt to fold the 5D MSSM with a single bulk Higgs hypermultiplet.  In fact, the MSSM version of that model predicts a Higgs boson with mass consistent with the now-observed value~\cite{Barbieri:2000vh}.  However, the corresponding prediction for the radius of compactification is $1/R \sim 700$ GeV, with a gluino around $350$~GeV.  One could hope that folding this model at arbitrary twist would yield a partial reduction of the negative, $y_t$-dependent Higgs soft mass, thereby raising $1/R$.  However, this fails because at non-trivial $\alpha$, the zero-mode Higgs doublet receives a positive KK mass at tree level.

One viable extension is to introduce a second bulk Higgs doublet.  An additional $SU(2)_H$ symmetry rotates the two Higgs hypermultiplets $H_{A,B}$ into each other.  The twisted boundary conditions can be modified to take advantage of this extra freedom in order to yield a single massless Higgs doublet at low energies.  Specifically, the action of the $\mathbb{Z}_2$ at $y=0$ is taken to be
\begin{align}
\left(
\begin{array}{cc}
H_A & H_B \\
H_A^{c\dag} & H_B^{c\dag}
\end{array}
\right)(-y)
&= 
\left[\sigma_3
\left(
\begin{array}{cc}
H_A & H_B \\
H_A^{c\dag} & H_B^{c\dag}
\end{array}
\right)\sigma_3\right](y),\notag\\[10pt]
\vspace{10pt}
\left(
\begin{array}{c}
\widetilde{H}_A \\
\widetilde{H}_B
\end{array}
\right)(-y)
&=
\sigma_3
\left(
\begin{array}{c}
\widetilde{H}_A \\
\widetilde{H}_B
\end{array}
\right)(y),\notag\\[10pt]
\left(
\begin{array}{c}
\widetilde{H}_A^c \\
\widetilde{H}_B^c
\end{array}
\right)(-y)
&=
-\sigma_3
\left(
\begin{array}{c}
\widetilde{H}_A^c \\
\widetilde{H}_B^c
\end{array}
\right)(y),
\end{align}
while the $2\,\pi\,R$ translation $\mathcal{T}$ is given by
\begin{align}
\left(
\begin{array}{cc}
H_A & H_B \\
H_A^{c\dag} & H_B^{c\dag}
\end{array}
\right)(y+2 \,\pi\,R)
&= 
\left[e^{-2\,\pi\,i\,\alpha\,\sigma_2}
\left(
\begin{array}{cc}
H_A & H_B \\
H_A^{c\dag} & H_B^{c\dag}
\end{array}
\right)e^{2\,\pi\,i\,\alpha\,\sigma_2}\right](y),\notag\\[10pt]
\vspace{10pt}
\left(
\begin{array}{c}
\widetilde{H}_A \\
\widetilde{H}_B
\end{array}
\right)(y+2 \,\pi\,R)
&=
e^{-2\,\pi\,i\,\alpha\,\sigma_2}
\left(
\begin{array}{c}
\widetilde{H}_A \\
\widetilde{H}_B
\end{array}
\right)(y),\notag\\[10pt]
\left(
\begin{array}{c}
\widetilde{H}_A^c \\
\widetilde{H}_B^c
\end{array}
\right)(y+2 \,\pi\,R)
&=
e^{-2\,\pi\,i\,\alpha\,\sigma_2}
\left(
\begin{array}{c}
\widetilde{H}_A^c \\
\widetilde{H}_B^c
\end{array}
\right)(y).
\end{align}
Note that we have implicitly set the $SU(2)_H$ twist angle equal to the $SU(2)_R$ twist angle.  Whether these angles are free parameters or instead determined by dynamics depends on physics above the cutoff.  If they were unequal, then the light Higgs boson would receive an $\mathcal{O}(1/R^2)$ tree-level positive mass squared proportional to the misalignment between these parameters.

This choice of boundary conditions leads to a single light Higgs scalar $h^{(0)}$ and heavier Higgs scalars and higgsinos with masses $\sim \alpha/R$; the KK decomposition is given in \eref{eq:BulkHiggsKKDef}.  One of the consequences of this configuration is the vanishing of the tree-level $D$-term quartic for $h^{(0)}$.  Therefore, in contrast to the brane Higgs model, an additional source of quartic is required in order to lift the Higgs boson mass to the observed value.  We choose to do this by adding a brane-localized singlet $S$ and an NMSSM-like coupling between the singlet and the Higgs bosons: 
\begin{align}
\mathcal{W} = \delta(y)\left(\lambda^{(5)}\,S \,H_A \, H^c_B + \frac{M_S}{2} S^2 + y_t^{(5)}\,H_A\,(Q\,U+Q_f\,U_f) \right).
\end{align}
Note that since the singlet has a large supersymmetric mass, it will not get an appreciable vev; furthermore, singlet-Higgs mixing will remain sufficiently small so as not to affect Higgs coupling measurements.

For the parameter space of interest, the singlet-Higgs coupling will be $\mathcal{O}(1)$ to accommodate $m_h = 125$ GeV.  In particular, the Higgs mass-squared is entirely determined by one-loop contributions, predominantly from the singlet and top Yukawa couplings (neglecting the small tree-level contribution from the singlet vev).  Therefore, it is important to include the leading radiative contributions to the Higgs potential from these couplings.  Following a similar justification to that given in Sec.~\ref{sec:BraneModel} above, the full Coleman-Weinberg potential from the (\f)top states is included.  The one-loop contribution to the Higgs mass parameter from loops involving the singlet is evaluated at fixed order, as there is a tree-level contribution to the quartic proportional to $\lambda^{(5)}$.  Note that there is no longer a tree-level quartic from the $SU(2)_W\times U(1)_Y$ gauge bosons.  However, the Higgs vev no longer breaks KK number/parity (although the top Yukawa coupling does), removing the proliferation of sums over KK modes in gauge/gaugino loops.  Thus we need only include the fixed-order contribution to the Higgs mass.  In analogy with \eref{eq:VBraneHiggsExpanded}, the potential for the zero-mode Higgs is
\begin{align}
V =&\,  (\lambda\,v^2-M_S\,v_S)^2 + 2\,\lambda^2\,v_S^2\,v^2 \notag \\[10pt]
&-\frac{9\, \Re\left[
\Li_5(e^{4\,i \,\pi (\alpha+\gamma)}) + \Li_5(e^{4\,i \,\pi (\alpha-\gamma)}) - 2\Li_5(e^{4\,i \,\pi \,\gamma})
\right]}{512\, \pi^6 \,R^4 }\notag\\[10pt]
&+\Big( \frac{3\,g^2 + g^{\prime 2}}{16\, \pi^4\, R^2} \Re\left[\zeta(3) - \Li_3(e^{2\,i\,\pi\, \alpha})\right] + m^2_{H,\,\rm Singlet} \Big) \, v^2,  
\end{align}
where $\vev{h^{(0)}} = v$, $\vev{S} = v_S$, $\gamma = \frac{1}{\pi} \tan^{-1}(\pi\, R \,y_t\, v)$,  $m^2_{H,\,\rm Singlet}$ is the soft mass contribution from the singlet sector (see \eref{eq:mHSinglet} in the appendix for the integral form),  and $\lambda =\lambda^{(5)}/( 4\, \pi\, R )$ is the rescaled singlet-Higgs coupling.  Note that $S$ acquires a tadpole that leads to a small vev $v_S = (M_S\, v^2\, \lambda)/(M_S^2 + 2 v^2 \lambda^2)$, which is included in the numerical results.

Now that we have the tools to explore the vacuum structure of the model, we may work out the dominant constraint on the parameter space.  Recall that in this model the gauginos, squarks, and Higgsinos are all bulk fields.  Given the choice of boundary conditions, the zero modes for all of these states receive $\alpha/R$-sized masses from boundary conditions.  Naively, this implies that the gluinos, squarks, and neutralinos will all be degenerate, while constraints on gluino-squark-neutralino simplified models will yield the strongest bounds.  To determine precise bounds, we must consider both the dominant gluino decay channels and the level of degeneracy between the various states.  The $\mathcal{O}(\alpha/R)$ estimate for the neutralino mass is modified by mixing due to the vev of the Higgs (along with the presence of the $S$-fermion, whose mass is similar to $\alpha/R$ in the viable parameter space).  The neutralino and chargino wave functions can be computed explicitly including the discontinuities that result from the Higgs and singlet vevs.  We find that the neutralino mass is degenerate with the colored superpartners with a splitting of at most $\mathcal{O}(20\%)$.\footnote{Note that the connection between degenerate spectra and 5D models was made previously in~\cite{Murayama:2012jh}, although there the Higgs was brane-localized so that the level of compression, which depended on the value of the $\mu$ parameter, was less robust.}

In addition to calculable contributions to the masses, there are incalculable contributions from brane-localized K\"{a}hler terms that can shift the mass spectrum as discussed in detail in Appendix~\ref{app:ZFactors}.  This can change the qualitative features of the LHC signatures in various ways.  While we expect the splitting between the gluino and lightest neutralino to be robust, for extreme cases this degeneracy could be spoiled.  Another interesting variation occurs by changing the identity of the lightest squark.  For example, if the stop becomes lighter than the light-flavor squarks, the gluino decay proceeds through one or two off-shell tops.  Gluino-squark associated production then involves a decay to a nearly degenerate neutralino, two tops, and light-flavor jets.  This specific signature has not been searched for directly at the LHC and presents an interesting opportunity for a new analysis.

While this somewhat-exotic signature is possible, for a wide range of the parameter space the dominant channel will be gluino and squark production with decays into light-flavor quarks and nearly-degenerate neutralinos.  Hence, jets + MET searches at the LHC are relevant~\cite{Aad:2014wea, Chatrchyan:2014lfa}.  An exclusion plot in the gluino/squark versus neutralino plane in the associated gluino-squark production channel for this model has been given in~\cite{Aad:2014wea}.  While this result does not include all the possible production modes in the simplified model, it does provide constraints on the largest production channel and therefore provides a reasonable approximation of the full limit.  In particular, this result captures the degree of degradation to the limits that results from degeneracy of the spectrum.  Therefore, we use this published limit as an estimate of the exclusion in our results below. 

\begin{figure}[t]
\centering
\includegraphics[width=.45 \textwidth]{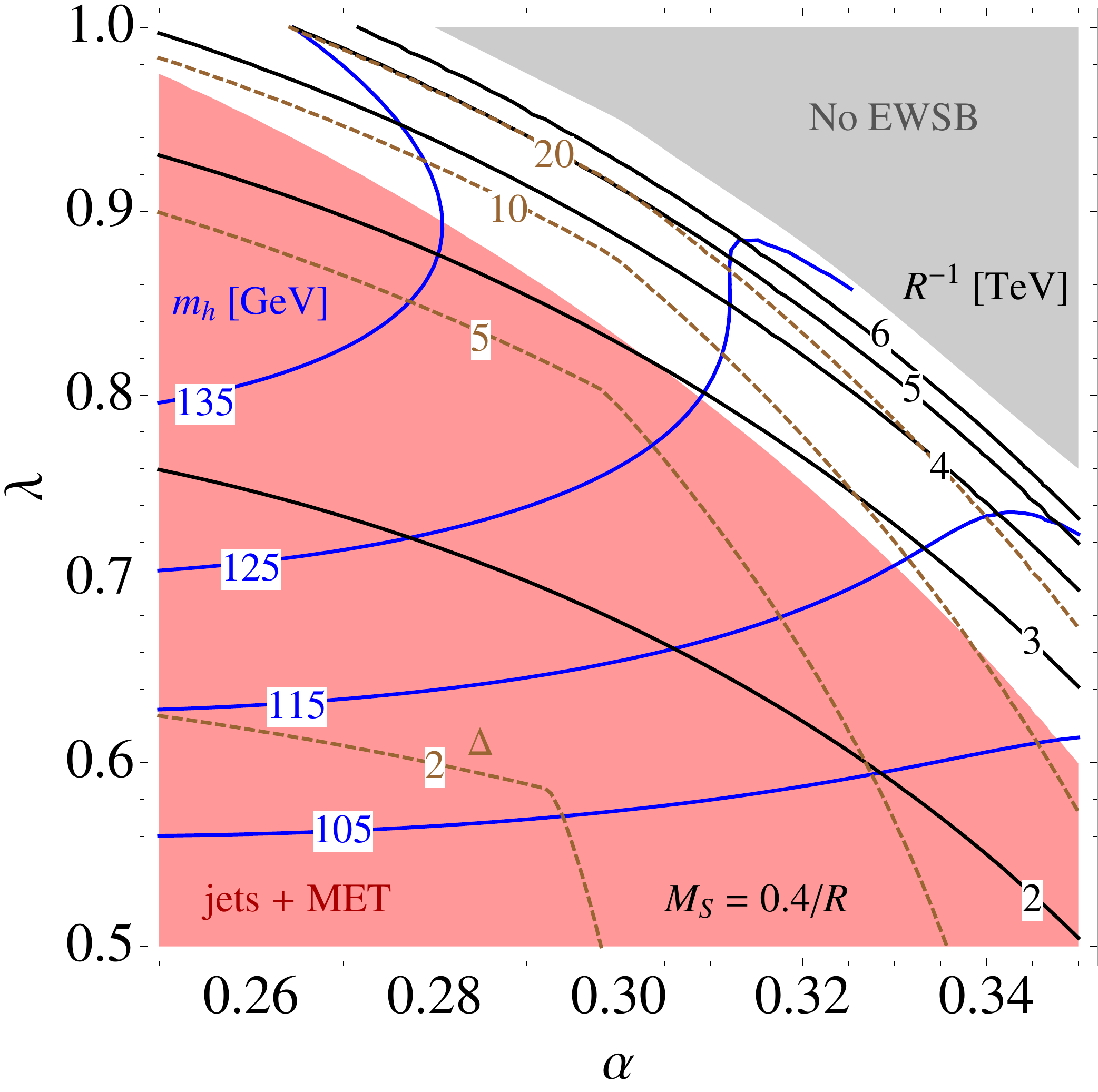} \quad \includegraphics[width=.45 \textwidth]{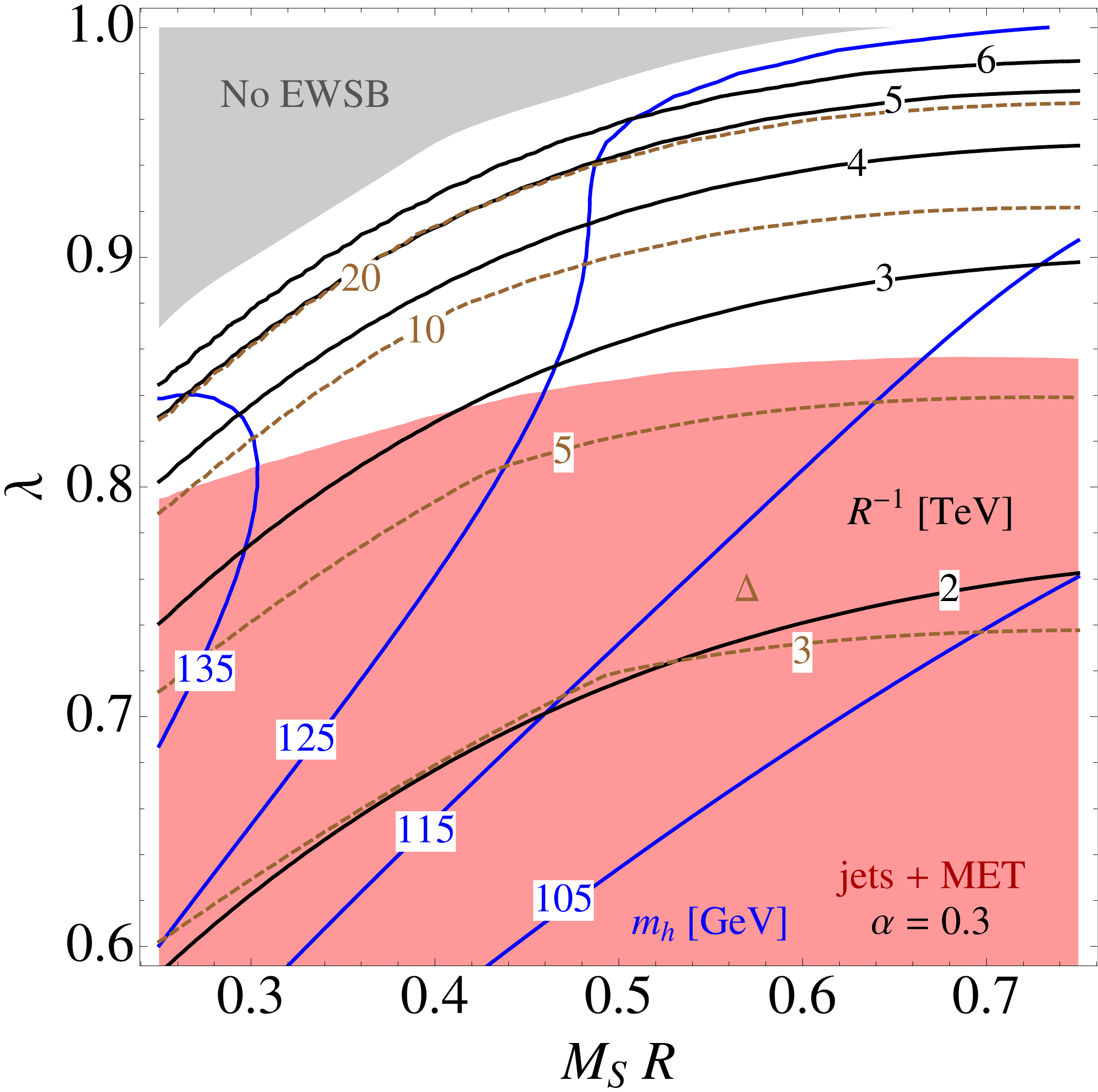}
\caption{Brane Higgs parameter space in the singlet-Higgs coupling $\lambda$ versus the twist $\alpha$ plane [left] and the $\lambda$ versus the singlet mass parameter $M_S$ plane [right].  The blue contours denote the mass of the Standard Model-like Higgs, while the black contours denote the inverse radius of compactification $1/R$.  Contours of fine-tuning are shown in brown.  The red shading indicates the exclusion coming from gluino mass limits relevant to the compressed parameter space of the model, while the grey shading indicates the region where EWSB does not occur.}
\label{fig:BulkHiggsParamSpace}
\end{figure}

Two slices of the viable parameter space are given in Fig.~\ref{fig:BulkHiggsParamSpace}: the left [right] panel shows $\lambda$ versus $\alpha$ [$M_S\,R$].  The red region is excluded from jets + MET searches at the LHC~\cite{Aad:2014wea, Chatrchyan:2014lfa}, and in the grey region there is no electroweak symmetry breaking.  Contours of the SM-like Higgs boson mass are shown in blue; the black contours denote $1/R$; and contours of fine-tuning\footnote{The fine-tuning measure is described as in the previous section with $X = \{y_t, g, \lambda, \alpha, M_S \}$.} are given in brown.  We can see that while the LHC has probed this parameter space, there is still a wide region where the tuning is mild ($\Delta \lesssim 10$).  Jets + MET searches at LHC13 will provide a strong test of the minimally tuned region of parameter space, and the observation of \f squarks would follow a discovery in this channel.  While the level of fine-tuning here is similar to that found in other effective theories of ``Natural SUSY'' with low cutoffs, it is worth emphasizing that \f SUSY provides a fully calculable model of SUSY breaking.  In comparison to other such models, {\it e.g.} low-scale gauge mediation, the fine-tuning in \f SUSY is quite mild.

\section{Conclusions}
\label{sec:Conc}

In this paper we have constructed the first calculable models of folded supersymmetry with successful electroweak symmetry breaking at one loop. To do so, we have generalized the original $f$-SUSY framework to incorporate non-trivial Scherk-Schwarz twist. We have also exploited some of the geographical freedom allowed by 5D theories, considering models in which the Higgs doublets are brane- or bulk-localized.

In brane-localized models, radiative corrections from bulk states naturally accommodate a 125 GeV Higgs. However, the smallness of certain soft-mass parameters in the Higgs sector leads to additional scalar and pseudoscalar Higgs bosons excluded by precision Higgs measurements. This may be addressed by deformations of the theory (such as the dimension-5 BMSSM operator) that raise the mass of the pseudoscalar Higgs, but even so Higgs coupling measurements provide the strongest constraint on the parameter space and push tuning of the weak scale to the sub-percent level.

In models where the Higgs doublets propagate in the 5D bulk, constraints from the Higgs sector are mitigated because the additional states are lifted by boundary conditions. However, the viable model introduced above has no $D$-term potential, requiring the introduction of new tree-level quartic couplings {\it a la} the NMSSM.  Although light folded states preserve the naturalness of the weak scale, the region of parameter space leading to electroweak symmetry breaking brings colored states into the reach of direct searches. The strongest bounds on these bulk Higgs models come from the gluino-squark-neutralino simplified model.  In light of current limits in the jets + MET channel, these models are tuned at better than $\sim 10 \%$.  

As a whole, this work highlights both the perils and the opportunities inherent in attempts at building concrete models of neutral naturalness.  Accommodating electroweak symmetry breaking and the observed Higgs mass pushes calculable realizations of \f SUSY away from the limit in which colored partner states are fully decoupled.  Rather, it favors regions with an admixture of conventional and folded phenomenology. The strongest bounds may come from more conventional channels, including precision Higgs measurements and colored sparticle searches.

Our work also highlights the vast range of possibilities contained within the \f SUSY paradigm.
There are a variety of worthwhile additional directions in which \f SUSY may be deformed while preserving the essential mechanism leading to neutral naturalness:

\begin{itemize}
\item {\it Further exploiting the geographical freedom in 5D models.} The original \f SUSY model featured brane-localized Higgses and bulk-localized matter fields. In this work we have considered the implications of bulk-localized fields, but in general there are a variety of possible configurations of bulk- and brane-localized fields that may have novel consequences.
\item {\it Exploring the opportunities provided by anomaly cancellation.} Crucially, the boundary conditions of \f SUSY leave no light folded fermions, so that zero-mode anomaly cancellation is insensitive to the precise choice of bulk folded states. Consistency of the 5D theory still may require the introduction of bulk Chern-Simons terms, but these do not significantly alter the folded mechanism. We have pointed out that this may be used to evade bounds on folded sleptons, but may be exploited more generally to produce more minimal models along these lines.
\item {\it Considering additional sources of SUSY breaking.} In this work we have restricted ourselves to SUSY breaking by boundary conditions in order to preserve the calculability of the Scherk-Schwarz twist. However, this restriction leads directly to many of the dominant constraints in our models, which may be ameliorated by the introduction of additional sources of SUSY breaking. In particular, this may imbue the original folded proposal with viable electroweak symmetry breaking and weaken bounds from conventional searches for SUSY partner states.
\end{itemize}

In the absence of new physics signals, the question of the smallness of the weak scale becomes increasingly pressing.  The exploration of novel theories that explain the lightness of the Higgs mass continues, and neutral naturalness represents a promising paradigm that accommodates null results but nonetheless yields experimentally observable consequences.  As we learn more from the now-ongoing run of the 13 TeV LHC, data will only sharpen the question and encourage further exploration.  This work has demonstrated that if we live in a folded world, then it is likely that we will know about it very soon. 

\section*{Acknowledgements}
We thank Nima Arkani-Hamed, Gustavo Burdman, Zackaria Chacko, David Curtin, Antonio Delgado, Daniel Harlow, Roni Harnik, Simon Knapen, Pietro Longhi, and Neal Weiner for useful conversations. TC is supported by an LHC Theory Initiative Postdoctoral Fellowship, under the National Science Foundation grant PHY-0969510.  NC is supported by the Department of Energy under the grant DE-SC0014129.

\appendix

\newcommand{\doublet}[2]
{
\left(
\begin{array}{c}
    #1 \\
    #2
\end{array}
\right)
}

\newcommand{\quadruplet}[4]
{
\left(
\begin{array}{c}
    #1 \\
    #2 \\
    #3 \\
    #4 
\end{array}
\right)
}

\section{Boundary Conditions} \label{app:bcs}
The starting point for all of our calculations will be to solve for the boundary conditions of the bulk fields. At arbitrary twist angle $\alpha$, doublets of $SU(2)_R$ obey the following transformation properties:
\begin{align}
\doublet{\Phi}{\Phi^{c \dag}}(y+ 2\,\pi\,R) &= 
(-1)^f e^{-2\,\pi\,i\,\alpha\,\sigma_2}
\left( \begin{array}{c}
\Phi \\
\Phi^{c \dag}
\end{array} \right) (y);
\notag \\
\doublet{\Phi}{\Phi^{c \dag}}(-y) &= 
\sigma_3
\doublet{\Phi}{\Phi^{c \dag}}(y)
\label{eq:app_boundary}
\end{align}
where $f=0$ (1) for the MSSM (folded) fields, and $\sigma_i$ are the Pauli matrices. Any generic $SU(2)_R$ twist can be brought to the form in Eq.~(\ref{eq:app_boundary}) through an $SU(2)_R$ transformation. 

For fields that do not transform under $SU(2)_R$, we define an ``effective'' Scherk-Schwarz twist. For example, the bulk matter fermions have the following transformation properties:
\begin{align}
\doublet{\psi}{\psi^c}(y+ 2\,\pi\,R) = 
(-1)^f 
\left( \begin{array}{c}
\psi \\
\psi^c
\end{array} \right) (y);
\qquad\qquad 
\doublet{\psi}{\psi^c}(-y) = 
\sigma_3
\doublet{\psi}{\psi^c}(y),
\label{eq:app_boundary_fermion}
\end{align}
where $(\psi,\psi^c)$ are the fermionic components of a bulk hypermultiplet. Since $e^{-i\,\pi\, f\, \sigma_2}=(-1)^f$, the MSSM (folded) fermions have an effective Scherk-Schwarz twist $\alpha=0$ ($1/2$). 

In order to simplify calculations involving both MSSM and folded bulk fields, the result can be obtained by assuming there is only one $SU(2)_R$ doublet with twist $\alpha$.  Then the contributions from the additional states is captured by replacing $\alpha$ with the appropriate effective Scherk-Schwarz twist listed in Table~\ref{table:app_effective_twist}.
\begin{table}[h!]
\renewcommand{\arraystretch}{2}
\setlength{\tabcolsep}{0.5em}
\centering 
\begin{tabular}{c|c|c|c|c}
&\multicolumn{2}{c|}{MSSM} & \multicolumn{2}{c}{Folded} \\
\hline
Fields & $(A^a_\mu, \Sigma^a), (Q, Q^c)$ & $(\lambda^a_1, i\lambda_2^a), (\wt Q, \wt Q^{c\dagger})$ &
$(Q_f, Q_f^c)$ & $(\wt Q_f, \wt Q_f^{c\dagger})$ \\
\hline
Effective SS-twist & $0$ & $\alpha$ & $\frac{1}{2}$ & $\alpha - \frac{1}{2}$ \\
\hline
\end{tabular}
\caption{The effective Scherk-Schwarz twist for different 5D bulk states. The results for a given field can be extrapolated from those for $SU(2)_R$ doublets by replacing $\alpha$ with the appropriate effective Scherk-Schwarz twist.  Note that for fields with zero effective SS-twist, an extra $1/\sqrt{2}$ factor must be included to appropriately normalize the 4D kinetic term.}
\label{table:app_effective_twist}
\end{table}

\pagebreak
\section{Wave functions} \label{app:wavefunctions}
Here we perform the KK expansions for all bulk fields. At non-maximal twist, squarks and gauginos mix with their $SU(2)_R$ partners.  Their KK expansions take the form of a sum over $SU(2)_R$ doublet wave functions, ${\bf \Psi}_n(y)$:
\begin{align}
\doublet{\!\!\!\wt Q}{\wt Q^{c\dagger}\!\!}
= \sum_{n=-\infty}^{\infty} Q^{(n)}\big(x^\mu\big)\,{\bf\Psi}_n(y)
\qquad 
\doublet{\,\,\,\lambda^a_1}{i\,\lambda_2^a}
= \sum_{n=-\infty}^{\infty} \lambda^{a}_{(n)}\big(x^\mu\big)\,{\bf\Psi}_n(y).
\end{align}
Here ${\bf \Psi}_n(y)$ needs to satisfy the boundary conditions in Eq.~(\ref{eq:app_boundary}), and is given by
\begin{align}
{\bf\Psi}_n(y) = 
\frac{1}{\sqrt{2\,\pi\, R}}
\doublet
{\cos\left( \frac{(n+\alpha)\,y}{R}\right)\vspace{5pt}}
{\sin\left( \frac{(n+\alpha)\,y}{R}\right)},
\label{eq:app_wavefunction}
\end{align}
where $n$ is an integer. The mass for each mode is ${m_n= |n+\alpha|/R}$.  The fermionic wave functions and their folded counterparts can be obtained via Table~\ref{table:app_effective_twist}. 

In general, Scherk-Schwarz symmetry breaking boundary conditions can involve symmetries other than $SU(2)_R$; this can lead to more complicated wave functions. For example, for our model where the Higgs fields are in the bulk, the Scherk-Schwarz twist can be embedded within any element of the $SU(2)_R \times SU(2)_H$ symmetry. One can fix an $SU(2)_R$ and $SU(2)_H$ basis such that the transformations for the Higgsinos are
\begin{align}
\left(
\begin{array}{cc}
\wt H_A & \wt H_A^c \\
\wt H_B & \wt H_B^{c}
\end{array}
\right)(y+2 \,\pi\,R)
&= 
\left[
e^{-2\,\pi\,i\,\beta\,\sigma_2}
\left(
\begin{array}{cc}
\wt H_A & \wt H_A^c \\
\wt H_B & \wt H_B^{c}
\end{array}
\right)\right](y);\notag \\[10pt]
\left(
\begin{array}{cc}
\wt H_A & \wt H_A^c \\
\wt H_B & \wt H_B^{c}
\end{array}
\right)(-y)
&= 
\left[
\sigma_3
\left(
\begin{array}{cc}
\wt H_A & \wt H_A^c \\
\wt H_B & \wt H_B^{c}
\end{array}
\right)
\sigma_3
\right](y).
\end{align}
and for the Higgses
\begin{align}
\left(
\begin{array}{cc}
H_A & H_B \\
H_A^{c\dag} & H_B^{c\dag}
\end{array}
\right)(y+2 \,\pi\,R)
&= 
\left[e^{-2\,\pi\,i\,\alpha\,\sigma_2}
\left(
\begin{array}{cc}
H_A & H_B \\
H_A^{c\dag} & H_B^{c\dag}
\end{array}
\right)e^{2\,\pi\,i\,\beta\,\sigma_2}\right](y),\notag\\[10pt]
\left(
\begin{array}{cc}
H_A & H_B \\
H_A^{c\dag} & H_B^{c\dag}
\end{array}
\right)(-y)
&= 
\left[
\sigma_3
\left(
\begin{array}{cc}
H_A & H_B \\
H_A^{c\dag} & H_B^{c\dag}
\end{array}
\right)
\sigma_3
\right](y).
\end{align}
The KK expansion for the Higgsinos are
\begin{align}
\doublet{\wt H_A}{\wt H_B} =  \sum_{n=-\infty}^{\infty} \wt H^{(n)}(x^\mu){\bf\Psi}_n(y);
\qquad 
\doublet{\wt H^c_A}{\wt H^c_B} =  \sum_{n=-\infty}^{\infty} \wt H^{c (n)}(x^\mu){\bf\Psi}_n(y),
\end{align}
where the wave functions ${\bf\Psi}_n(y)$ are the same as those for the squarks given above in \eref{eq:app_wavefunction} with $\alpha \rightarrow \beta$.
The $\big(\wt H^{(n)}, \wt H^{c (n)}\big)$ pair up to form a tower of Dirac fermions with masses $|n+\beta|/R$. For the Higgs bosons, the KK expansion mixes all four scalars. Ignoring electroweak symmetry breaking, we find
\begin{align}
\left(
\begin{array}{cc}
H_A & H_B \\
H_A^{c\dag} & H_B^{c\dag}
\end{array}
\right) = 
\sum_{n=-\infty}^{\infty} \bigg[
h^{(n)} (x^\mu)
\, {\bf h}_n(y)+ H^{(n)} (x^\mu)\, {\bf H}_n(y)
\bigg].
\label{eq:BulkHiggsKKDef}
\end{align}
where the matrices ${\bf h}_n(y)$ and ${\bf H}_n(y)$ are given by
\begin{align}
{\bf h}_n(y)
&=
\frac{1}{\sqrt{4\,\pi\, R}}
\left(
\begin{array}{cc}
\,\,\,\,\,\cos\left(\frac{(n+\alpha-\beta)\,y}{R}\right) & \sin\left(\frac{(n+\alpha-\beta)\,y}{R}\right) \\[10pt]
-\sin\left(\frac{(n+\alpha-\beta)\,y}{R}\right) & \cos\left(\frac{(n+\alpha-\beta)\,y}{R}\right)
\end{array}
\right);
\notag \\[10pt]
{\bf H}_n(y)
&=
\frac{1}{\sqrt{4\,\pi\, R}}
\left(
\begin{array}{cc}
\,\,\,\,\,\cos\left(\frac{(n+\alpha+\beta)\,y}{R}\right) & -\sin\left(\frac{(n+\alpha+\beta)\,y}{R}\right) \\[10pt]
-\sin\left(\frac{(n+\alpha+\beta)\,y}{R}\right) & -\cos\left(\frac{(n+\alpha+\beta)\,y}{R}\right)
\end{array}
\right).
\end{align}
Before electroweak symmetry breaking, the masses for the $h^{(n)}$ and $H^{(n)}$ modes are $|n+\alpha -\beta|/R$ and $|n+\alpha+\beta|/R$, respectively. Fixing $\alpha=\beta$ results in a massless $h^{(0)}$ which is desirable as a starting point for models of natural electroweak symmetry breaking. From the 4D perspective, $h^{(0)}$ is the SM-like Higgs field that develops an electroweak symmetry breaking vev $v$. This leads to additional $v$-dependent contributions to the mass spectrum not explicitly shown here.  Note that because $H_A$ and $H_B^{c\dag}$ contribute equally to $h^{(0)}$, its potential will not receive contributions from the gauge $D$-terms.

\section{Boundary Fields and Discontinuities}
\label{app:bdy_field_discont}
When the 5D Lagrangian contains interaction terms localized on the boundaries, the 5D wave functions can develop kinks and discontinuities. One such important interaction comes from the Higgs; a non-zero vev for a 4D Higgs field (either a boundary or KK state) $\langle H \rangle = v$ induces a brane-localized mass.  Consider the superpotential for the top sector at finite $v$,
\begin{align}
\mathcal{W} \supset  -\delta(y)\big(2\,\pi\, R\, y_t \,v \big)\, U\, T + U^c\, \partial_5 U + T^c \,\partial_5 T,
\end{align}
where $y_t$ is the 4D top-yukawa coupling. The same brane-localized interaction exists for the brane Higgs and the bulk Higgs scenarios ($v$ is replaced by $v_u$ in the brane Higgs case). To avoid any ambiguities when integrating across the 5$^\mathrm{th}$ dimension due to the $\delta(y)$ function, we will work with the interval $(-\epsilon, 2\,\pi\, R - \epsilon)$, where $\epsilon >0$ is small.  A Higgs vev induces mixing between all four fields $\big(\,\wt{t}, \wt{u}^c, \wt{u}^\dagger, \wt{t}^{c\dagger}\big)$, and one expects their KK expansions to be a sum of vectors involving all four fields,
\begin{align}
\quadruplet{\!\!\wt{t}}{\wt{u}^c}{\wt{u}^\dagger}{\,\wt{t}^{c\dagger}}
= \sum_{n=-\infty}^{\infty}\bigg[
\,\wt t^+_{(n)}\big(x^\mu\big)\, {\bf T}_n^+(y) +
\wt t^-_{(n)}\big(x^\mu\big)\, {\bf T}_n^-(y)
\bigg].
\end{align}
Here ${\bf T}_n^\pm(y)$ are the 5D wave functions that satisfy the equation of motion in the domain $y\in(-\epsilon, 2\,\pi\, R-\epsilon)$: 
\begin{align}
\big(\partial_5 + \Delta^\dagger\big)\big(\partial_5 - \Delta\big){\bf T}^\pm=-\left(m^\pm_n\right)^2 {\bf T}^\pm;
\qquad
\Delta=
-\big(2\,\pi\, R\,y_t\, v\big)\delta(y)
\setlength\arraycolsep{6pt}
\left(\;
\begin{matrix}
0 & 0 & 0 & 0\\
1 & 0 & 0 & 0\\
0 & 0 & 0 & 0\\
0 & 0 & 1 & 0
\end{matrix}
\;
\right).
\label{eq:app_kink_eom}
\end{align}
The $\delta(y)$ terms in $\Delta$ will force ${\bf T}$ to develop a discontinuity at $y=0$. Due to the Scherk-Schwarz boundary conditions, the discontinuity is extended to all integer multiples of $2\,\pi\, R$. Away from the discontinuities, the equation of motion are still solved by the cosine and sine functions in Eq.~(\ref{eq:app_wavefunction}), up to $n$-dependent phase shifts. In order to introduce a periodic discontinuity to the wave function, it is convenient to introduce the function $\{y\} \equiv y \text{ mod } (2\,\pi\, R)$; $\{y\}$ maps the real line onto the fundamental domain $[0, 2\,\pi\, R)$ and contains a finite jump from $2\,\pi\, R$ to 0 at every integer multiple of $2\,\pi\, R$. 

Doublets of $SU(2)_R$ that satisfy the boundary conditions and have a discontinuity at $y=0$ can be written as
\begin{align}
{\bf \Psi}_{k}(y)=\frac{1}{\sqrt{2\,\pi\, R}}
\doublet{\cos\left( \frac{k\,\{y\}+\alpha \,y}{R}-\pi \,k\right)}
{\sin\left( \frac{k\,\{y\}+\alpha\, y}{R} - \pi\, k\right)}
\qquad
\{y\} \equiv y\text{ mod } 2\,\pi\, R ,
\label{eq:app_discont_wave}
\end{align}
where $k$ is not restricted to integer values.  The extra $-\pi\, k$ phase ensures that $\sigma_3 {\bf \Psi}_{k}(-y)={\bf \Psi}_{k}(y)$. Similar wave functions have been considered before in~\cite{Bagger:2001qi}.

\begin{figure}[t!]
\includegraphics[width=.45 \textwidth]{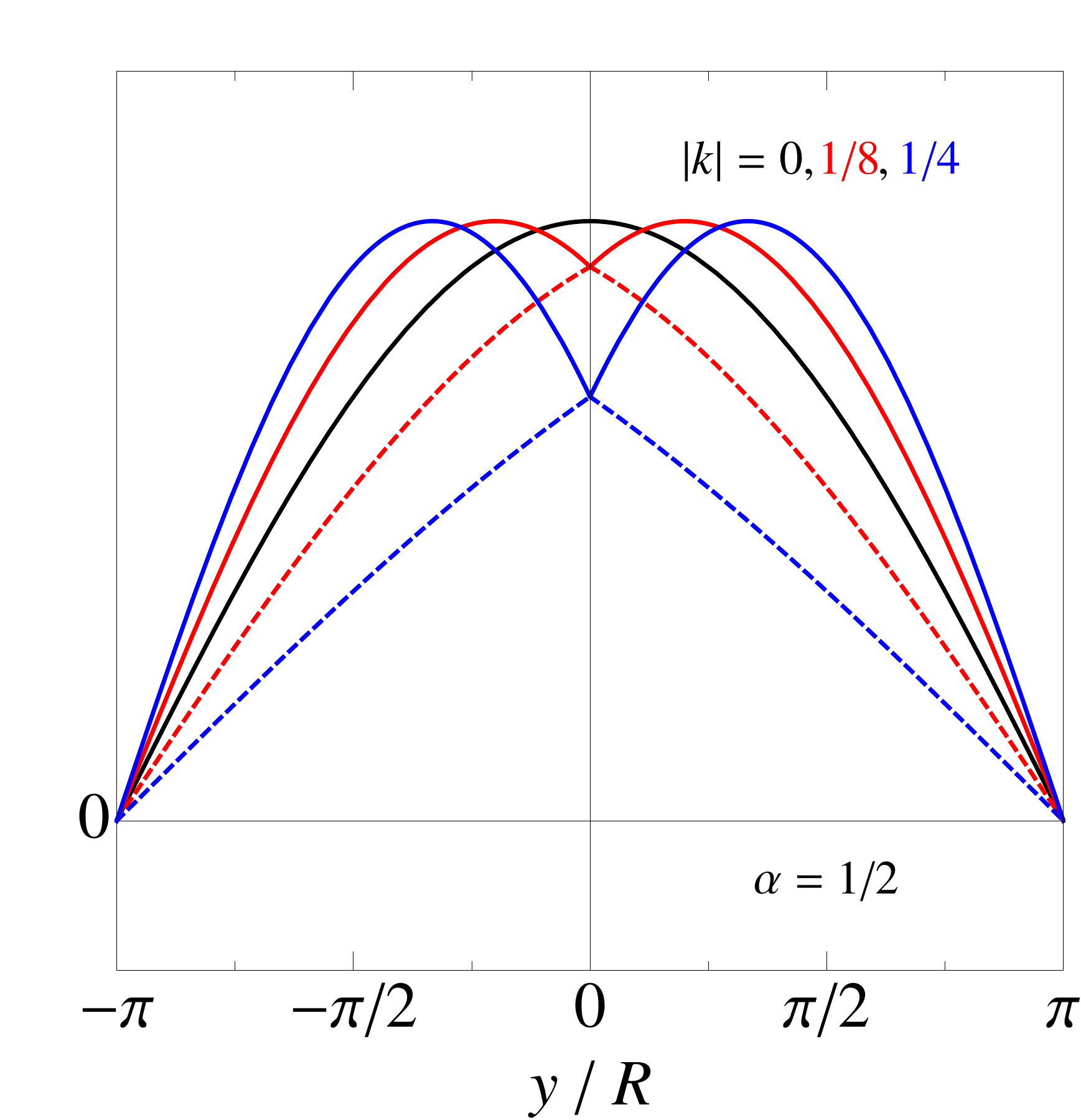} \quad \includegraphics[width=.45 \textwidth]{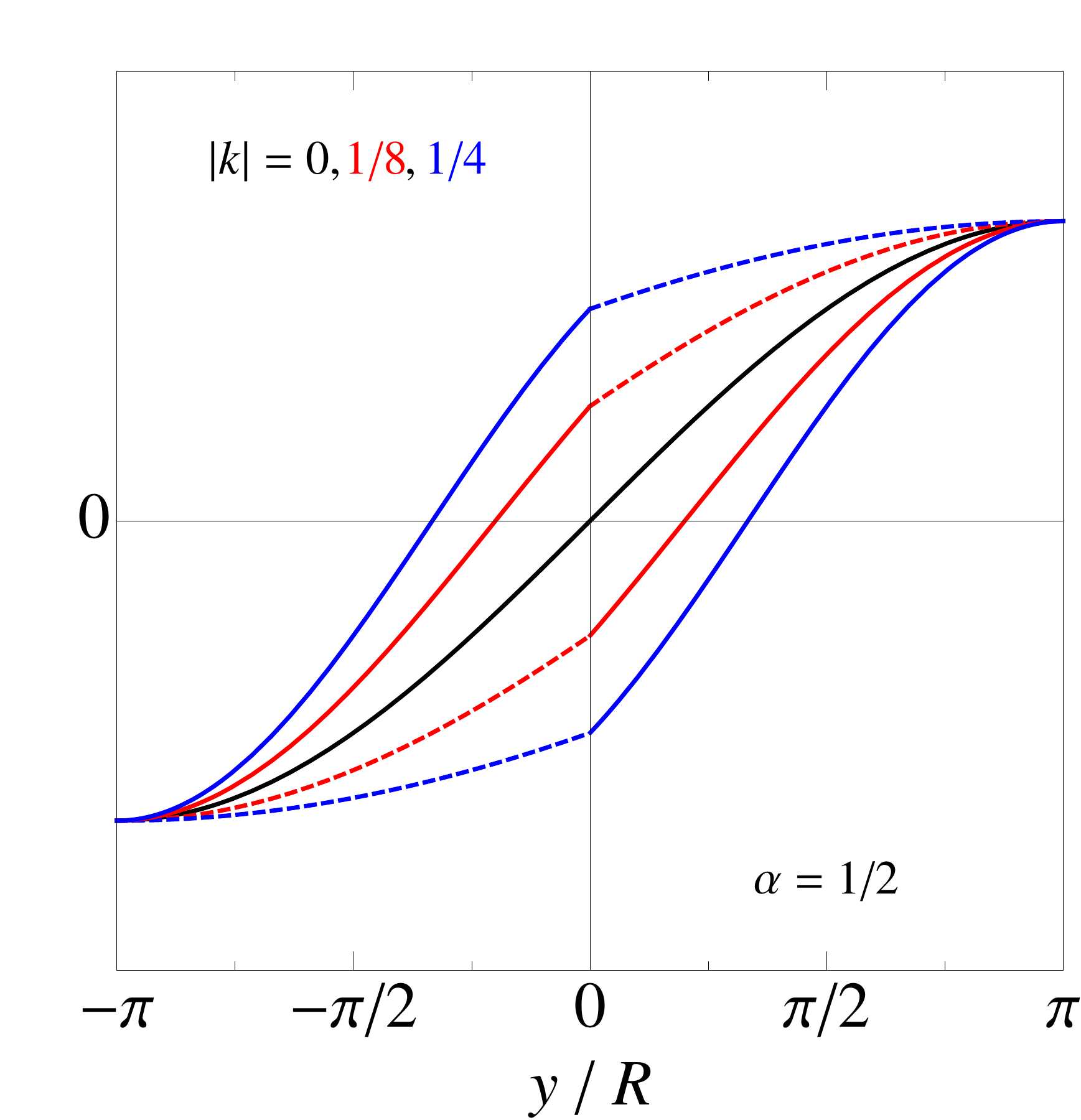}
\begin{minipage}{0cm}
\vspace*{-14.4cm}\hspace*{-22.8cm}{Even}
\vspace*{0.5cm}
\end{minipage}
\begin{minipage}{0cm}
\vspace*{-14.4cm}\hspace*{-7.2cm}{Odd}
\vspace*{0.5cm}
\end{minipage}
\caption{Sample discontinuous wave functions ${\bf\Psi}_{k}(y)$ shown in arbitrary units as a function of $y/R$, for $\alpha = 1/2$.  The left (right) panel shows the shape for even (odd) boundary conditions at the $y = 0$ brane.  The black, red, and blue lines denote ${\bf\Psi}_{k}(y)$ for $|k| = 0,1/8,$ and $1/4$, respectively.  The solid (dashed) lines correspond to positive (negative) $k$.}
\label{fig:app_discont_wavefunction}
\end{figure}

Note that ${\bf \Psi}_{k}(-y)$ matches the continuous version in Eq.~(\ref{eq:app_wavefunction}) when $k$ is taken to be an integer. A sketch of the two components of the wave function ${\bf\Psi}_{k}(y)$ with $\alpha = 1/2\,$ is shown in Fig.~\ref{fig:app_discont_wavefunction}. The cosine component develops a kink but remains continuous since it is even under reflection about $y=0$. On the other hand, the sine component develops a finite, discontinuous jump across $y=0$ and is odd under reflection about $y=0$.

Taking the derivative for the sine component, the discontinuity at $y = 0$ gives a delta function.  In order to satisfy the equation of motion, the $\delta(y)$ terms from $\partial_5$ and $\Delta$ must cancel, which leads to an equation for $k$ that has discrete solutions $k_n$. The KK modes are found to be
\begin{align}
{\bf T}_n^\pm=
\frac{1}{\sqrt{4\,\pi\, R}}
\begin{pmatrix}
\cos\left( \frac{k_n\,\{y\}+\alpha \, y}{R}-\pi\, k_n\right) \\
\pm\sin\left( \frac{k_n\,\{y\}+\alpha\,y}{R}-\pi \,k_n\right)\\
\pm\cos\left( \frac{k_n\,\{y\}+\alpha\,y}{R}-\pi \,k_n\right) \\
\sin\left( \frac{k_n\,\{y\}+\alpha \,y}{R}-\pi \,k_n\right)
\end{pmatrix}
\qquad 
k_n= n \pm \frac{1}{\pi}\tan^{-1}\big(\pi \,R\,y_t \,v \big),
\label{eq:app_topwave}
\end{align}
where $n$ is any integer. The operator $\partial_5 - \Delta$ acting on ${\bf T}$ simply picks out the smooth part of the derivative; thus the masses of the KK modes are given by ${m_n^\pm = |n + \alpha \pm \tan^{-1}(\pi\, R\,y_t\, v)/\pi|/R}$. 

\section{One-loop Coleman-Weinberg Potential}\label{app:Potential}
In order to evaluate the vacuum structure of the models discussed in the main text, the Coleman-Weinberg contribution to the Higgs potential from the top/\f top sectors must be evaluated. At one loop~\cite{Coleman:1973jx}
\begin{align}
V_{\text{CW}}(v) = \frac{1}{2}\sum_{i\in \rm DOF}(-1)^F\, {\rm Tr}\log
\Big(\partial^2 + m_i^2(v) \Big),
\label{eq:app_CW}
\end{align}
where $F=0$ ($1$) for scalars (fermions), and $v$ is the Higgs vev. In principle, the summation is over all degrees of freedom that couple to the Higgs, but here we will focus on the contributions proportional to $y_t$. If the states running in the loop descend from bulk fields, a summation over KK modes is required.  For the (\f)top sector, the $v$ dependent KK masses in \eref{eq:app_topwave} are analytic, and the one-loop Coleman Weinberg potential can be computed analytically:
\begin{align}
V_{y_t}  = 
\frac{6}{R^4}\int_0^{\frac{\tan^{ \text{\tiny -1}\! } (\pi\, R\, y_t \,v)}{\pi}} 
\!\!\!\!\!\!\!\!
\text{d}\eta \int \!\frac{\text{d}^4 \ell}{(2\,\pi)^4}
\sum_{n=-\infty}^{\infty}
\bigg[
&
\frac{\frac{n}{2}+\alpha+\eta}{(\frac{n}{2}+\alpha+\eta)^2+\ell^2} -
\frac{\frac{n}{2}+\eta}{(\frac{n}{2}+\eta)^2+\ell^2} - \notag\\
&
\frac{\frac{n}{2}+\alpha-\eta}{(\frac{n}{2}+\alpha-\eta)^2+\ell^2} +
\frac{\frac{n}{2}-\eta}{(\frac{n}{2}-\eta)^2+\ell^2}
\bigg],
\end{align}
where we have combined the contributions from the top and $f$-top sector together under the same summation over $n$ for convenience.  The potential can then be evaluated as sums of polylogarithms,
\begin{align}
V_{y_t} =& -\frac{9 \Re\left[
\Li_5(e^{4\,i \,\pi \,(\alpha+\gamma)}) + \Li_5(e^{4\,i \,\pi\, (\alpha-\gamma)}) - 2\,\Li_5(e^{4\,i\, \pi \,\gamma})
\right]
}{512\, \pi^6 \,R^4 },\notag\\
\gamma \equiv&\, \frac{1}{\pi} \tan^{-1}(\pi \,R\, y_t\, v),
\end{align}
where constant terms have been dropped.
Note that at maximal twist, $V_{y_t}$ becomes $v$-independent (at one loop) due to the bifold protection.

\section{One-loop Fixed-order Calculations}
\label{app:fixed_order}
In addition to the Coleman-Weinberg potential from the top sector, there are other important one-loop contributions to the Higgs potential which are discussed in the following subsections.

\subsection{Gauge soft mass}

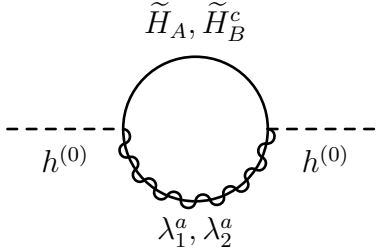
\begin{figure}[h!]
\vspace{0.5cm}
\centering
    \begin{fmffile}{oneloopgauge}
      \setlength{\unitlength}{1cm}
      \begin{fmfgraph*}(5,3.5)
        \fmfleft{i}
        \fmfright{o}
        \fmf{dashes,tension=1.5,label=$h^{(0)}$}{i,v1}
        \fmf{vanilla,left,tension=0.4, label=$\wt H_A,, \wt H_B^c$}{v1,v2}
        \fmf{vanilla,right,tension=0.4}{v1,v2}
        \fmfset{wiggly_slope}{85}
        \fmf{photon,right,tension=0.4, label=$\lambda^a_1,, \lambda^a_2$}{v1,v2}
        \fmf{dashes,tension=1.5,label=$h^{(0)}$}{v2,o}
      \end{fmfgraph*}
    \end{fmffile}
\caption{Gaugino contributions to the zero-mode (bulk) Higgs soft mass. Other contributions from gauge interactions and $D$-term quartics can be obtained from this diagram from the limit $\alpha \rightarrow 0$.  Analogous diagrams contribute to the soft mass in the case of the brane-localized Higgs.}
\label{fig:app_softmass_gauge}
\end{figure}

The bulk gauginos obtain tree-level, SUSY-breaking masses from boundary conditions.  These in turn yield a contribution to the Higgs soft mass at one loop. For convenience, we explicitly compute only the gaugino loops (see Fig.~\ref{fig:app_softmass_gauge}); the contribution from the bosonic gauge sector can be extracted by the replacement $\alpha \rightarrow 0$ in the gaugino result.  

When the Higgs is brane-localized, it only couples to the $\mathcal{N}=1$ vector multiplet that is even under reflection about the $y=0$ brane, such that only the diagram involving $(\lambda^a_1, \wt h)$ is present. For the bulk Higgs case, there are multiple diagrams involving different combinations of gauginos and Higgsinos.

A straightforward diagrammatic computation reveals that the gauge soft mass in the brane and bulk Higgs models are the same (ignoring possible, small corrections $\sim (\mu R)^2$):
\begin{align}
m^2_{H,\,\rm gauge} &= \frac{3\,g^2 + g^{\prime 2}}{R^2}
\int\frac{\text{d}^4 \ell}{(2\,\pi)^4}
 \sum_{n=-\infty}^{\infty} \left(\frac{1}{n^2+\ell^2} - \frac{1}{(n+\alpha)^2+\ell^2}\right) \notag \\
&= \frac{3\,g^2 + g^{\prime 2}}{16\, \pi^4 \,R^2} \Re\Big[
\zeta(3) -  \Li_3(e^{2\,i\,\pi\, \alpha})
\Big].
\end{align}
Unlike the bulk matter contributions, the gauge soft mass is largest at maximal twist since there is no bi-fold protection in the gauge sector.

\subsection{$b_\mu$}
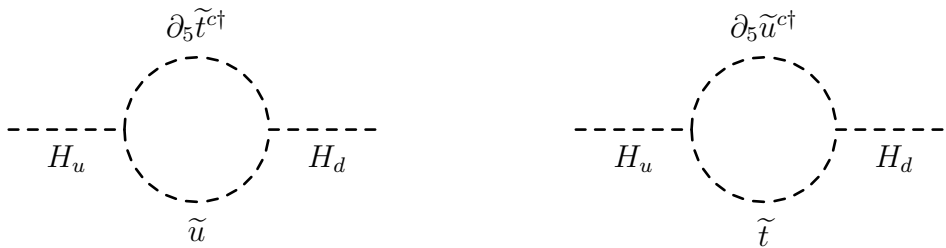
\begin{figure}[h!]
\vspace{0.5cm}
\centering
\begin{minipage}{.45\textwidth}
\centering
    \begin{fmffile}{oneloopbmu1}
      \setlength{\unitlength}{1cm}
      \begin{fmfgraph*}(5,2.5)
        \fmfleft{i}
        \fmfright{o}
        \fmf{dashes,tension=1,label=$H_u$}{i,v1}
        \fmf{dashes,left,tension=0.4, label=$\partial_5 \wt t^{c\dagger}$}{v1,v2}
        \fmf{dashes,right,tension=0.4, label=$\wt u$}{v1,v2}
        \fmf{dashes,tension=1,label=$H_d$}{v2,o}
      \end{fmfgraph*}
    \end{fmffile}
\end{minipage}
\begin{minipage}{.45\textwidth}
\centering
    \begin{fmffile}{oneloopbmu2}
      \setlength{\unitlength}{1cm}
      \begin{fmfgraph*}(5,2.5)
        \fmfleft{i}
        \fmfright{o}
        \fmf{dashes,tension=1,label=$H_u$}{i,v1}
        \fmf{dashes,left,tension=0.4, label=$\partial_5 \wt u^{c\dagger}$}{v1,v2}
        \fmf{dashes,right,tension=0.4, label=$\wt t$}{v1,v2}
        \fmf{dashes,tension=1,label=$H_d$}{v2,o}
      \end{fmfgraph*}
    \end{fmffile}
\end{minipage}
\vspace{5mm}
\caption{One-loop contributions to $b_\mu$ from the stop sector.  Analogous diagrams involving the \f stops are also included.  The gauge contribution is numerically subdominant and not shown here.}
\label{fig:app_bmu}
\end{figure}
The brane Higgs model presented in Sec.~\ref{sec:BraneModel} of the main text relies on a $\mu$ term in the brane-localized superpotential in order to give the Higgsinos non-zero mass.  Since the tree-level spectrum exhibits SUSY breaking, a $b_\mu$ term is generated at one loop.  There are two contributions to $b_\mu$, one proportional to the top Yukawa and another from gauge interactions. The (\f)squark contribution relies on the following interactions from the $F$-term potential:
\begin{align}
\mathcal{L} \supset
\int \text{d}^5 x \, \delta(y)\left[
y_t^{(5)} \, H_u \Big( 
\,\wt t \,\partial_5 \wt u^{c \dagger}-
\wt u \, \partial_5 \wt t^{c\dagger}
\Big)
- \mu \,y_t^{(5)} \, H_d \, \wt t^\dagger \, \wt u^\dagger
 + \textrm{h.c.}
\right].
\end{align}
Figure~\ref{fig:app_bmu} shows the one-loop contribution to the $b_\mu$ term from the top/stop sector. Resolving the $\delta(y)$ in these interactions requires a double summation over the KK tower -- the resulting contribution dominates over that from gauge loops, since the latter only contains a single sum. Explicitly,
\begin{align}
b_\mu &= \frac{3\,y_t^2\, \mu}{R} 
\int \frac{\text{d}^4 \ell}{(2\,\pi)^4}
\sum_{n_1,n_2}
\left[
\frac{n_1 + n_2 + 2\alpha }{
\left[(n_1 + \alpha)^2 + \ell^2\right]
\left[(n_2 + \alpha)^2 + \ell^2\right]
}
+
\left\{ \alpha \rightarrow \alpha - \frac{1}{2} \right\}
\right] \notag \\
&=\frac{3\, y_t^2\, \mu}{16\, \pi^3\, R}  \Im\big[\Li_2\big(e^{4\,i\,\pi\, \alpha}\big)\big].
\end{align}
From this result, it is straightforward to see that $b_\mu$ vanishes at $\alpha=0,1/2$, since in both limits the different components of an $SU(2)_R$ doublet do not mix, \emph{i.e.}, a $U(1)_R$ symmetry is preserved. At $\alpha = 1/4$, $b_\mu$ is also zero as a result of an accidental cancellation between the top and \f-top sector; the dominant contribution to $b_\mu$ comes from the gauge sector for this choice of $\alpha$. As discussed in Sec.~\ref{sec:BraneModel} above, the parameter space near $\alpha=1/4$ is not phenomenologically allowed due to large Higgs mixing.  We are therefore justified in ignoring the gauge contribution to $b_\mu$ in the viable regions of parameter space.

Naively, the convergence of this loop expansion is not particularly well-behaved as one includes additional insertions of the Higgs vevs.  Operationally, the brane-localized top Yukawa comes with a $\delta(y)$, yielding a proliferation of KK sums.  This contribution should, in principle, be resummed via a Coleman-Weinberg calculation in the 2D space of vevs $(v_u, v_d)$. However, the parameter space that results in natural electroweak symmetry breaking requires that $\mu \ll 1/R$; higher order effects involving $\mu\, v_u $ are expected to be subdominant.  Furthermore, in the viable model presented in the main text, there is a tree-level contribution to an effective $b_\mu$ which dominates the physics of electroweak symmetry breaking.  Therefore, we conclude that this level of approximation is sufficient for the exploration of the parameter space discussed in Sec.~\ref{sec:BraneModel}. 

\subsection{Singlet Soft Mass}
\begin{figure}[h!]
\vspace{0.5cm}
\centering
\begin{minipage}{.32\textwidth}
\centering
    \begin{fmffile}{singlet1}
      \setlength{\unitlength}{0.8cm}
      \begin{fmfgraph*}(5,2.5)
        \fmfleft{i}
        \fmfright{o}
        \fmf{dashes,tension=1,label=$h^{(0)}$}{i,v1}
        \fmf{dashes,left,tension=0.4, label=$S$}{v1,v2}
        \fmf{dashes,right,tension=0.4, label=$h^{(n)},,H^{(n)}$}{v1,v2}
        \fmf{dashes,tension=1,label=$h^{(0)}$}{v2,o}
      \end{fmfgraph*}
    \end{fmffile}
\end{minipage}
\begin{minipage}{.32\textwidth}
\centering
    \begin{fmffile}{singlet2}
      \setlength{\unitlength}{0.8cm}
      \begin{fmfgraph*}(5,1.5)
        \fmftop{t}
        \fmfleft{i,it}
        \fmfright{o,ot}
        \fmf{dashes,tension=1,label=$h^{(0)}$}{i,v2}
        \fmf{dashes,tension=1,label=$h^{(0)}$}{v2,o}
        \fmffreeze
        \fmf{dashes,right,tension=0.2}{v2,t,v2}
        \fmflabel{$S, h^{(n)},H^{(n)}$}{t}
        \fmf{phantom}{it,ot}
      \end{fmfgraph*}
    \end{fmffile}
\end{minipage}
\begin{minipage}{.32\textwidth}
\centering
    \begin{fmffile}{singlet3}
      \setlength{\unitlength}{0.8cm}
      \begin{fmfgraph*}(5,2.5)
        \fmfleft{i}
        \fmfright{o}
        \fmf{dashes,tension=1,label=$h^{(0)}$}{i,v1}
        \fmf{vanilla,left,tension=0.4, label=$\wt S$}{v1,v2}
        \fmf{vanilla,right,tension=0.4, label=$\wt h^{(n)},, \wt H^{(n)}$}{v1,v2}
        \fmf{dashes,tension=1,label=$h^{(0)}$}{v2,o}
      \end{fmfgraph*}
    \end{fmffile}
\end{minipage}
\vspace{5mm}
\caption{ One-loop contributions to the Higgs soft mass from the singlet sector in the bulk Higgs scenario.}
\label{fig:app_softmass_singlet}
\end{figure}
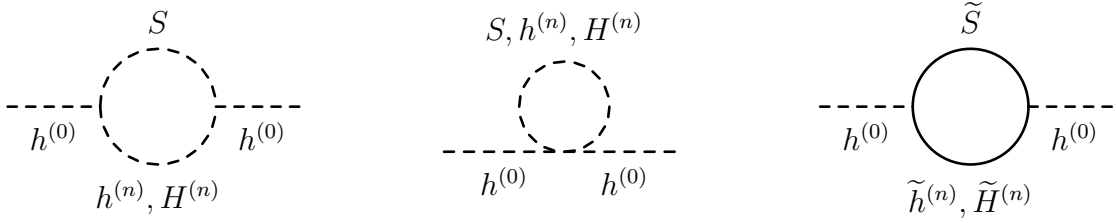
In the bulk Higgs model, a brane-localized singlet $S$ with an NMSSM-like coupling to the Higgs states is included in order to reproduce the measured value of the physical Higgs mass.  In the parameter space of interest, this coupling is large enough that the leading loop correction involving this interaction should be included.

The superpotential for this model includes
\begin{align}
\mathcal{W} \supset \delta(y)\left( \lambda^{(5)}\, S\, H_A \, H_B^c + \frac{M_S}{2} \,S^2 \right) + H_A^c \,\partial_5 H_A + H_B^c \,\partial_5 H_B,
\end{align}
which can be used to construct the one-loop singlet-Higgs diagrams shown in Fig.~\ref{fig:app_softmass_singlet}. Explicitly, this yields
\begin{align}
m^2_{H,\,\rm Singlet} = \frac{\lambda^2}{2\,\pi\, R^2}\int_0^\infty
\frac{\text{d}\ell \, \ell^2}{M_S^2\,R^2 + \ell^2}
&\bigg[
\frac{4\,\ell^2\,\sin^2(\pi\, \alpha)\coth(\pi\, \ell)- 2\,\ell\,M_S \, R \,\sin(2\,\pi\, \alpha)}{\cosh(2\,\pi\, \ell)-\cos(2\,\pi\, \alpha)}\, + \notag \\
&  \frac{2 \,M_S^2 \,R^2 \sin^2(2\,\pi\, \alpha)\coth(\pi\, \ell)+ \ell\,M_S\,R\,\sin(4\,\pi\, \alpha)}{\cosh(2\,\pi\, \ell)-\cos(4\,\pi\, \alpha)}
\bigg],
\label{eq:mHSinglet}
\end{align}
where we have defined $\lambda \equiv \lambda^{(5)}/( 4\, \pi\, R )$ as the effective 4D coupling. The integral is evaluated numerically for the results provided in the text.  Since the lightest states in the singlet/Higgs sector are bosons (in contrast to the top/stop sector) $m^2_{H,\,\rm Singlet}$ is positive. For the parameter space of interest, the positive soft mass from the singlet coupling is smaller in magnitude than the soft mass from the top sector.

\section{Brane Localized Kinetic Terms}
\label{app:ZFactors}
The 5D theories of interest become strongly coupled at scales $\Lambda$ not too far above $1/R$.  One way that this non-perturbative physics manifests is in the form of incalculable brane-localized K\"ahler potentials,
\begin{align}
\mathcal{K}\supset  \delta(y)\sum_i Z_{\mathcal{O}_i} \mathcal{O}_i
 + \delta(y-\pi \,R)\sum_j Z'_{\mathcal{O}_j} \mathcal{O'}_j,
\end{align}
where $\mathcal{O}_i$ ($\mathcal{O'}_j$) are K\"ahler operators that respect the local $\mathcal{N}=1$ supersymmetry at $y=0$ ($\pi \,R$).  Ignoring gauge interactions, the lowest-dimension operators are of the form $\delta(y)\, Z_{ij}\,\Phi_i^\dagger \,\Phi_j  + \delta(y-\pi\, R)\, Z'_{ij}\,\Phi_i'^\dagger\, \Phi'_j $, which could, in principle, lead to sizable corrections to the Higgs potential. 
Note that, in general, these arbitrary brane-localized kinetic terms lead to large rates for flavor violating observables.  We will assume that the flavor structure of these terms is controlled by MFV\footnote{The $Q^c$ superfield is required to have the conjugate flavor charges of $Q$.} \cite{D'Ambrosio:2002ex}, such that we can neglect these constraints. 

It is also possible that large $Z$-factors which descend from strong dynamics could spoil the feature of calculability espoused above.  Using naive dimensional analysis (NDA), the sizes of these $Z$-factors are estimated to be $\sim 3\,\pi/(2\,\Lambda) \sim R$ \cite{Chacko:1999hg}. In this section, we will study their impact with a focus on corrections to the Higgs potential.

Of particular concern are the $Z$-factors at $y=\pi R$.  The $\mathbb{Z}_2^f$ symmetry that exchanges matter superfields with their folded counterparts is not respected by boundary conditions on this brane; as a result, no relationship between matter and folded $Z$-factors should be assumed.  The brane-localized $Z$-factors kink the wave functions for the relevant states, modifying the Higgs/(\f)squark wave function overlap at $y = 0$ and splitting the effective top Yukawa couplings between the MSSM and folded sectors.  This could spoil the bi-fold protection of the KK spectrum, potentially leading to large corrections to the Higgs potential.  Nevertheless, the underlying supersymmetry of the UV theory prevents the re-introduction of quadratic divergences.

In practice, the discontinuities derived from the $Z$-factors lead to transcendental expressions for the KK spectrum, spoiling summability of the KK contributions in terms of simple analytic functions. To keep numerical calculations tractable, we only consider in detail the effects of adding a $Z$-factor for the $U$ type fields at the $y=\pi\, R$ brane, corresponding to adding the following terms to the K\"ahler potential: 
\begin{align}
\mathcal{K}\, \supset Z'_U\,\delta(y-\pi \,R)\abs{U'}^2 + Z'_{U_f} \,\delta(y-\pi\, R)\abs{U'_f}^2.
\end{align}
Here the prime on $U'$ denotes the superfield whose scalar component is even around the $y=\pi \, R$ brane. In the following, we will only focus on the scalars as the fermionic wave functions can be extrapolated from the scalar ones by taking $\alpha \rightarrow 0$. The equations of motion for $\wt u'$, the scalar component of $U'$, are
\begin{align}
\big(1+Z'_U \,\delta(y-\pi \,R)\big)\partial^2  \wt u' - \partial_5^2 \wt u' = 0 
\qquad
\partial^2 \wt u^{c\prime } - 
\partial_5\left[\frac{1}{1+Z'_U\,\delta(y-\pi \,R)} \,\partial_5 \wt u^{c\prime } \right]= 0 .
\label{eq:app_Zfactor_eom}
\end{align}
The $\delta(y-\pi\, R)$ function will introduce discontinuities in the 5D wave functions. The ansatz from Eq.~(\ref{eq:app_discont_wave}) applies, and the resulting KK expansion is given by
\begin{align}
\binom{\!\wt u'}{\,\,\wt u^{c\prime \dagger}} =
\sum_{n} \,\wt u'_{(n)}(x^\mu)\, \frac{{\bf \Psi}_{k_n}(y-\pi \,R)}{\sqrt{1+\frac{Z'_U\cos^2(\pi\, k_n)}{2\,\pi\, R}}}\,,
\quad \text{with} \quad
-\frac{Z'_U(k_n+\alpha)}{2\,R} = \tan(\pi \,k_n).
\label{eq:app_Z_mass}
\end{align}

\begin{figure}[t!]
\includegraphics[width=.6 \textwidth]{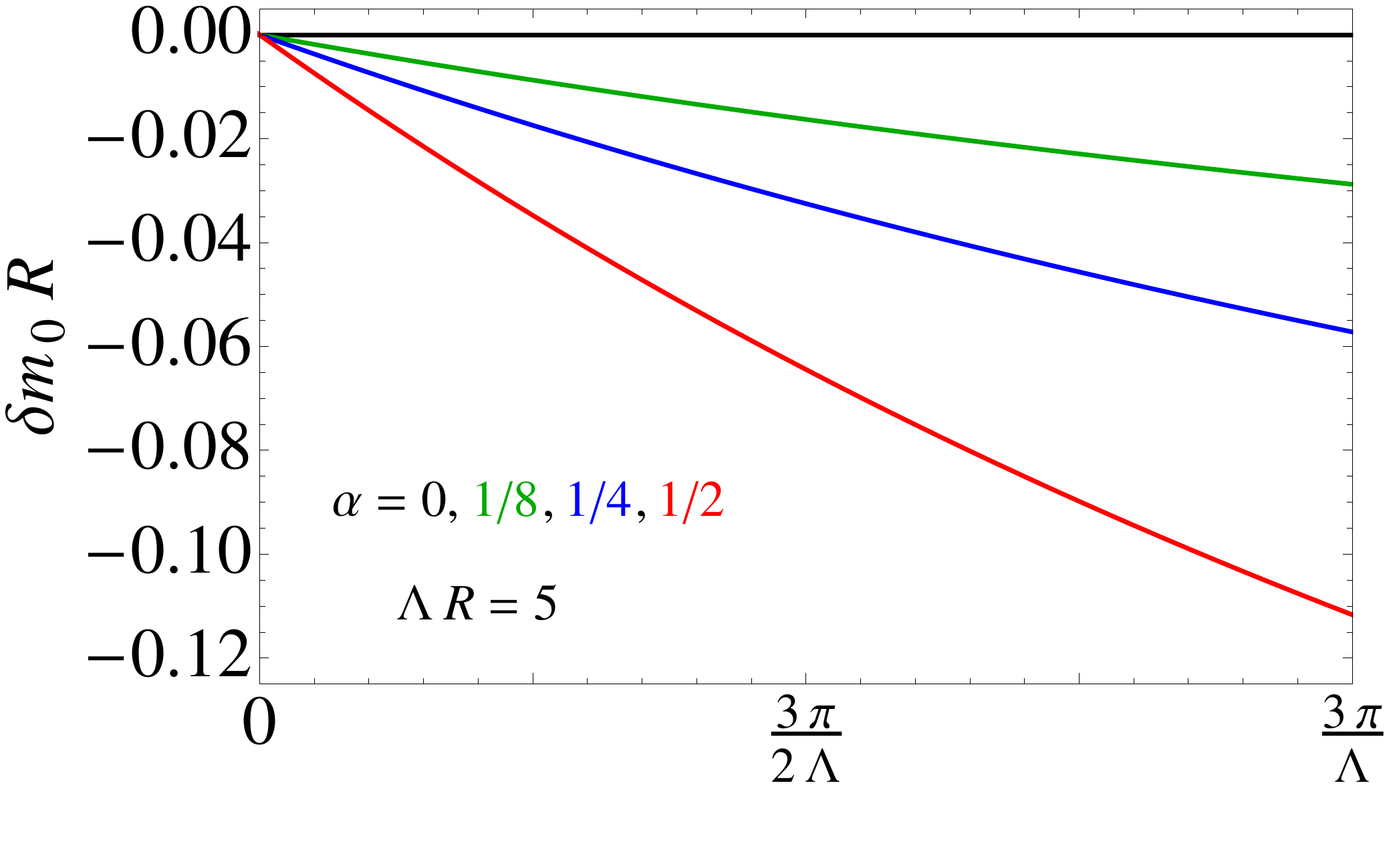}
\begin{minipage}{0cm}
\vspace*{0cm}\hspace*{-8.3cm}{\large$Z'_{U}$}
\end{minipage}
\caption{The mass shift $\delta m_0$ of the $n=0$ KK mode in units of $1/R$ as a function of $Z'_{U}$ for $\alpha=0, 1/8, 1/4, 1/2$. Here we have taken $\Lambda \,R = 5$. At $\alpha=0$, $\delta m_{0}$ is identically zero by supersymmetry.}
\label{fig:app_Zfactor_deltam0}
\end{figure}
The mass of each KK mode is given by $m_n = |k_n + \alpha|/R$. Figure~\ref{fig:app_Zfactor_deltam0} shows the change in $m_0$ as a function of $Z'_U$. At $\alpha=0$, the mass shift is exactly zero due to the restoration of accidental SUSY.  

A positive $Z$-factor of order $3\,\pi/2\,\Lambda$ from NDA yields a 10-20\% shift in the mass of the lightest KK mode. The mass-shifts for higher KK modes are smaller. As a result of these shifts,  the spectrum of bulk matter fields can only be reliably computed at the 10-20\% level.  This implies an inherent uncertainty in the level of compression between the colored superpartners and the LSP (and in the identity of the LSP itself, in some cases).  

The Higgs couplings to the top/\f top sectors also depend on the $Z$-factors. In fact, the 4D Higgs-squark and Higgs-quark coupling are not equal due to the different 5D wave functions at $y=0$.  This mismatch of couplings would naively lead to hard SUSY breaking and reintroduce quadratic divergences, as the loop contributions do not cancel level by level. However, this viewpoint mistakenly assumes a common 4D cutoff for scalars and fermions, which violates 5D Lorentz invariance. A better approach to study the UV behavior is to perform the calculations in terms of winding modes in the 5$^{\rm th}$ dimension~\cite{Cheng:2002iz}, where all divergences come from integrals over 5D momenta.  It can be shown explicitly that no quadratic sensitivity to the cutoff appears in the calculation of the Higgs soft mass due as expected to the presence of 5D SUSY.
  
\begin{figure}[t!]
\vspace{0.5cm}
\centering
\begin{minipage}{.45\textwidth}
\centering
    \begin{fmffile}{oneloopmass1}
      \setlength{\unitlength}{1cm}
      \begin{fmfgraph*}(5,2.5)
        \fmfleft{i}
        \fmfright{o}
        \fmf{dashes,tension=1,label=$H$}{i,v1}
        \fmf{dashes,left,tension=0.4, label=$\partial_5 \wt t^{c\dagger},,\partial_5 \wt u^{c\dagger}$}{v1,v2}
        \fmf{dashes,right,tension=0.4, label=$\wt u,,\wt t$}{v1,v2}
        \fmf{dashes,tension=1,label=$H$}{v2,o}
      \end{fmfgraph*}
    \end{fmffile}
\end{minipage}
\begin{minipage}{.45\textwidth}
\centering
    \begin{fmffile}{oneloopmass2}
      \setlength{\unitlength}{1cm}
      \begin{fmfgraph*}(5,1.5)
        \fmftop{t}
        \fmfleft{i,it}
        \fmfright{o,ot}
        \fmf{dashes,tension=1,label=$H$}{i,v2}
        \fmf{dashes,tension=1,label=$H$}{v2,o}
        \fmffreeze
        \fmf{dashes,right,tension=0.2}{v2,t,v2}
        \fmflabel{$\wt u,\wt t$}{t}
        \fmf{phantom}{it,ot}
      \end{fmfgraph*}
    \end{fmffile}
\end{minipage}
\vspace{5mm}
\caption{One-loop contributions to the Higgs soft mass from squark loops in the presence of brane-localized kinetic terms. The fermionic diagrams are not shown; their contributions can be extrapolated from the scalar ones by substituting $\alpha \rightarrow 0$.}
\label{fig:app_softmass_top}
\end{figure}
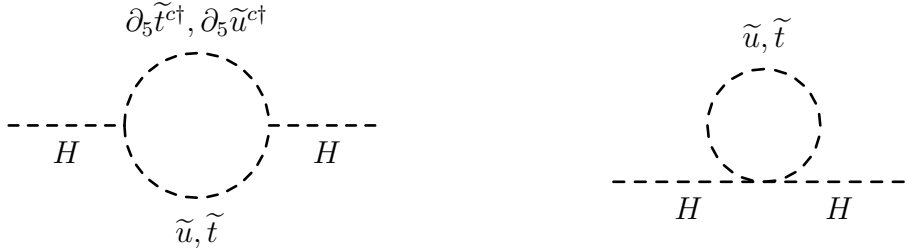

Next we will use the KK decomposition in Eq.~(\ref{eq:app_Z_mass}) to compute the modifications to the Higgs potential. Transcendentality of the KK spectrum implies that an analytic Coleman-Weinberg calculation is not possible. Hence, a fixed-order calculation will be presented to provide an estimate for the effects of $Z$-factors. 

First, consider the correction to the one-loop Higgs soft mass. The squark contributions come from the two sets of diagrams shown in Fig.~\ref{fig:app_softmass_top}. The quark contributions can be extrapolated from these results by substituting $\alpha\rightarrow 0$.  The full one-loop Higgs soft mass is found to be
\begin{align}
m^2_{H, y_t}
= \frac{2\,y_t^2}{R^2}
\int \frac{\text{d}^4 \ell}{(2\,\pi)^4} \sum_{n_1, n_2}
&\bigg\{
\bigg[
\frac{\ell^2\, \psi_{n_2}(0)^2 + 2\,(n_1 +\alpha)(k_{n_2} +\alpha)\,\psi_{n_2}(0)}
{
\big[(n_1+\alpha)^2 + \ell^2\big]
\big[(k_{n_2}+\alpha)^2 + \ell^2\big]
} - 
 \big(\alpha \rightarrow 0 \big)
\bigg] \notag \\
&\qquad + \bigg[ \textrm{folded sector} \bigg] \bigg\},
\label{eq:app_Z_softmass}
\end{align}
where $\psi_{n_2}(0)$ is the wave function factor at the $y=0$ brane and is given by $1/\sqrt{1+Z'_U\,\cos^2(\pi \,k_{n_2})/(2\,\pi\, R)}$. The contributions from the folded sector can be extrapolated from the MSSM result by substituting $\alpha \rightarrow \alpha - 1/2$ and $Z'_U \rightarrow Z'_{U_f}$. To obtain explicit results, we truncate the series and perform a numerical integration over the remaining spectrum. The integration bound and the truncation order is fixed such that the numerical answer is within $\lesssim 1\%$ of the analytic answer when $Z'_U = Z'_{U_f}=0$. 

\begin{figure}[t!]
\centering
\quad\quad\includegraphics[width=.4\textwidth]{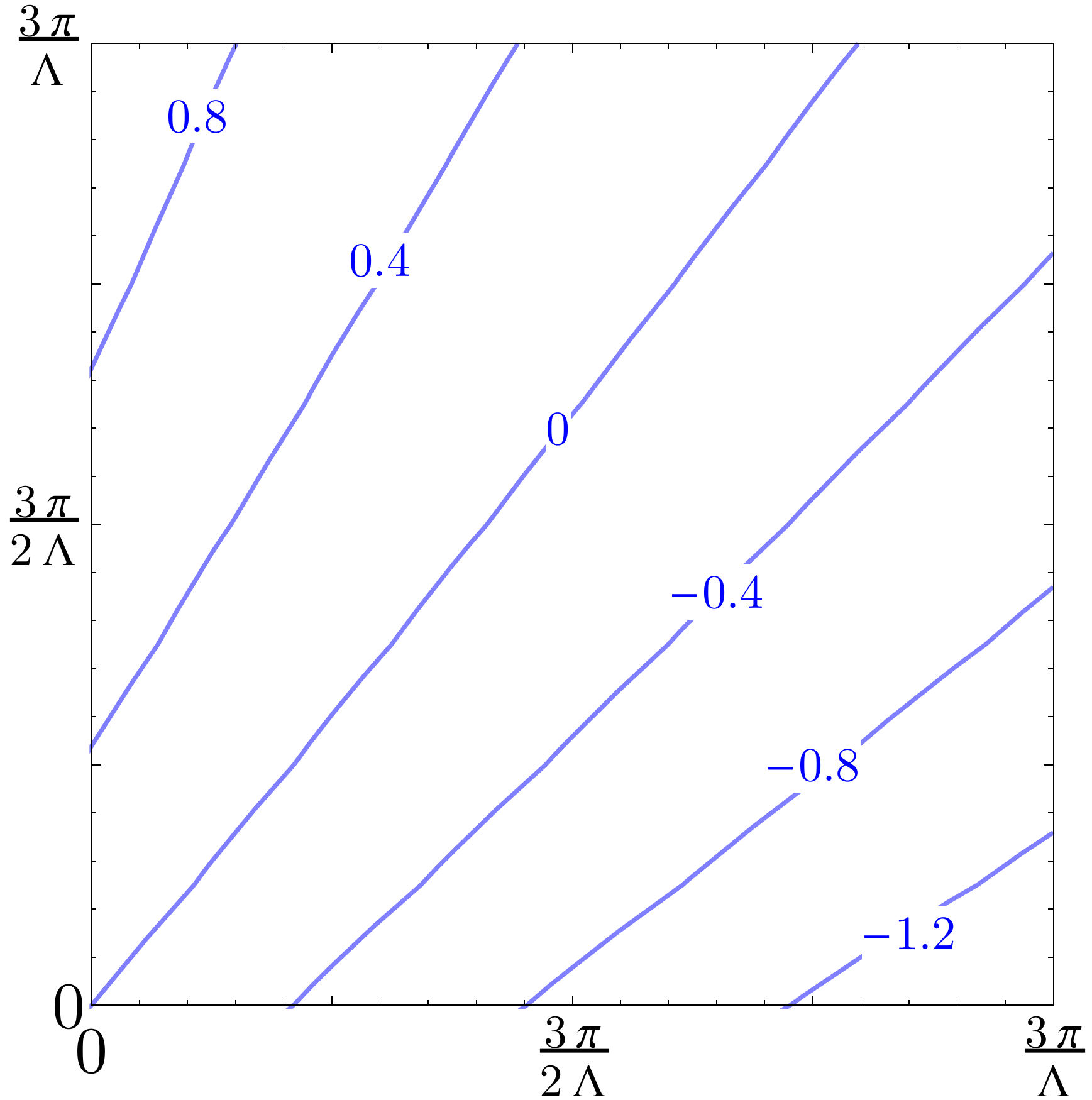}\quad\quad\quad\quad \includegraphics[width=.4\textwidth]{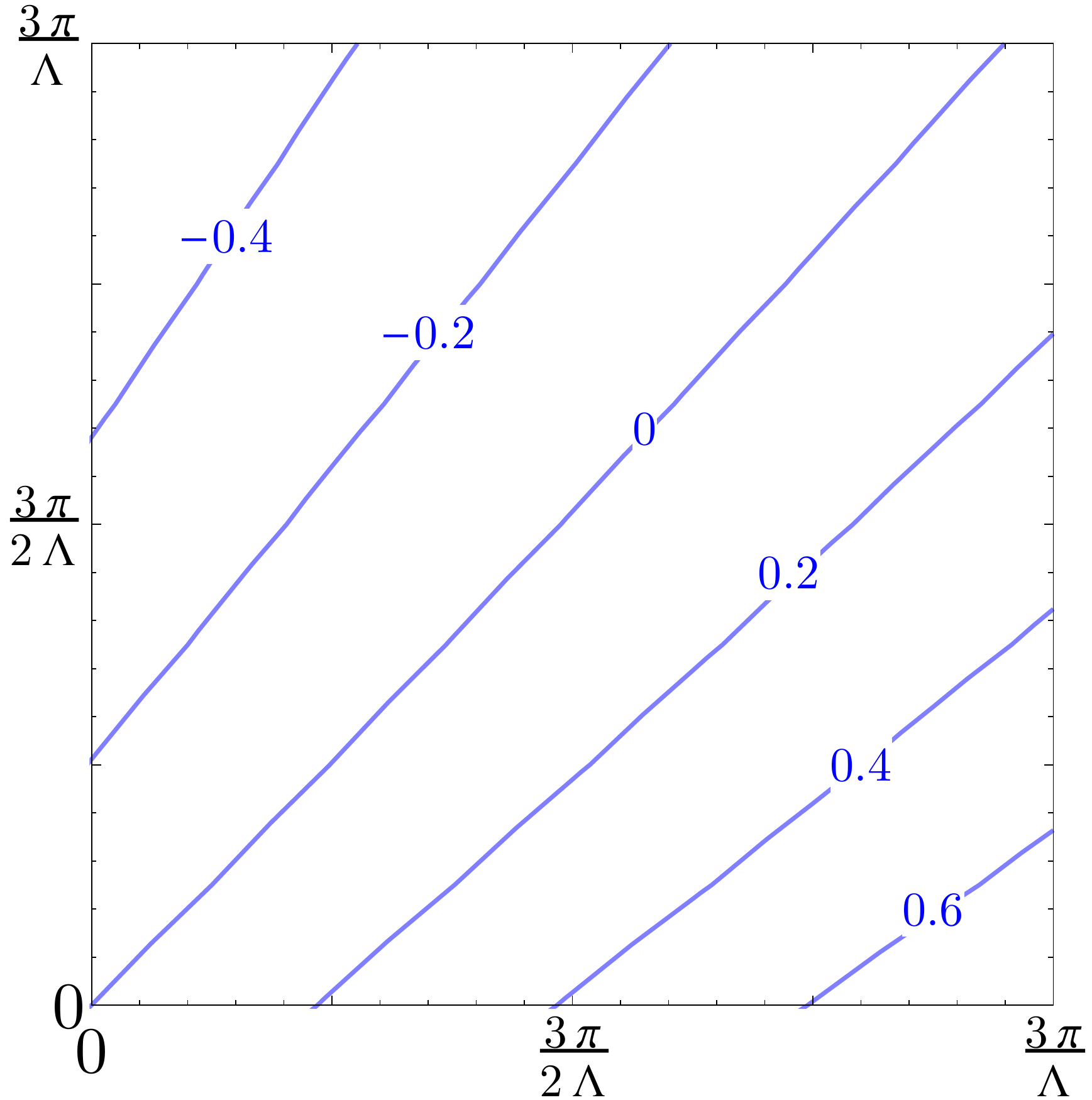}
\begin{minipage}{0cm}
\vspace*{-6.2cm}\hspace*{-30.6cm}\rotatebox{90}{{$Z'_{U_f}$}}
\vspace*{0.5cm}
\end{minipage}
\begin{minipage}{0cm}
\vspace*{1.7cm}\hspace*{-23cm}{$Z'_{U}$}
\vspace*{0.5cm}
\end{minipage}
\begin{minipage}{0cm}
\vspace*{-6.2cm}\hspace*{-14.6cm}\rotatebox{90}{{$Z'_{U_f}$}}
\vspace*{0.5cm}
\end{minipage}
\begin{minipage}{0cm}
\vspace*{1.7cm}\hspace*{-7cm}{$Z'_{U}$}
\vspace*{0.5cm}
\end{minipage}
\caption{Fractional shifts in the Higgs soft mass (left) and $b_\mu$ term (right) due to brane-localized kinetic terms $Z'_U, Z'_{U_f}$ at $\alpha = 1/3$. For the purpose of the NDA estimate, we take $\Lambda\,R = 5$.  The fractional shifts depend only weakly on $\alpha$, save at $\alpha=0,1/2$ where the shifts are identically zero and the fractional changes are ill-defined (also at $\alpha=1/4$, where the $b_\mu$ from the top and \f-top cancels accidentally at one-loop). }
\label{fig:app_Zfactor_deltam}
\end{figure}

The left panel of Fig.~\ref{fig:app_Zfactor_deltam} shows the fractional change of the Higgs soft mass from the top sector as a function of $(Z'_U, Z'_{U_f})$ at $\alpha=1/4$. The fractional changes are not very sensitive to the choice of $\alpha$. For comparable $Z'_U \simeq Z'_{U_f} > 0$, a $\lesssim 30$\% change in the soft mass is observed.  Corrections to the Higgs mass parameter from wave function factors on the $y = 0$ brane can also be read off of the figure.  Noting that at $y=0$ the $\mathbb{Z}_2^f$ symmetry forces the MSSM and folded $Z$-factors to be equal, it is clear that the corrections from these contributions are only $\mathcal{O}(10\%)$.

It is also a logical possibility that the $Z$-factors are negative.  However, in this case the transcendental relationship for $k_n$ given in \eref{eq:app_Z_mass} has a complex solution.  We will now demonstrate that this leads to the appearance of a zero-norm state in the KK spectrum.\footnote{In this respect we disagree with the statement that has been made previously in the literature that $Z<0$ can lead to tachyonic or negative norm states at tree level, \emph{e.g.}~\cite{delAguila:2003bh}.}  As a simple example, take a generic bulk hypermultiplet $(\Phi,\Phi^c)$ with a non-zero brane kinetic term for $\Phi$ on the $y = 0$ brane.  Assuming the $\phi$ $(\phi^c)$ scalar is even (odd) on the $y=0$ brane, the presence of the brane kinetic term will induce a kinked wave function:
\begin{align}
\left(
\begin{array}{c}
\phi\\
\phi^c
\end{array}
\right)= \frac{1}{\sqrt{2\,\pi\,R}}\sum_n A_n 
\left(
\begin{array}{c}
\cos\Big(\frac{k_n\,\{y\}+\alpha\,y}{R}-\pi\,k_n\Big)\\
\sin\Big(\frac{k_n\,\{y\}+\alpha\,y}{R}-\pi\,k_n\Big)
\end{array}
\right).
\end{align}
Then keeping track of the complex value of $k_n$ the normalization is given by
\begin{align}
A_n = \left[\frac{\sinh (2\,\pi\,\text{Im}(k_n))}{2\,\pi\,\text{Im}(k_n)}+\frac{Z}{4\,\pi\,R}\big[\cos(2\,\pi\,\text{Re}(k_n))+\cosh(2\,\pi\,\text{Im}(k_n))\big]\right]^{-1/2}.
\end{align}
Using the transcendental relation that defines $k_n$, it is straightforward to show that the only consistent solution is $A_n = 0$ for the complex $k_n$.  Therefore, the spectrum contains a state with zero norm at tree level.

The presence of this state may indicate that negative $Z$-factors could not be generated by well-behaved UV theories.  For example, this state should receive a kinetic term at one loop; in principle, this can be negative, violating unitarity.  In any case, the computation underlying Fig.~\ref{fig:app_Zfactor_deltam} does not obviously generalize to include states with zero norm.  We leave a full investigation of the impact of negative brane-localized $Z$-factors to future work. 

We have checked that if one excludes this zero-norm state, NDA-sized negative $Z$-factors can lead to $\mathcal{O}(1)$ corrections to the Higgs mass parameter.  Thus the assumption of calculability, which is central to the arguments made in the main text, may only hold for positive $Z$-factors.  However, it is unclear whether or not the contribution of the neglected zero-norm states would yield a compensating contribution to the Higgs soft mass.

A similar computation can be carried out for the $b_\mu$-term. The one-loop diagrams are given in Fig.~\ref{fig:app_bmu}, with the result
\begin{align}
b_\mu 
= \frac{3\,y_t^2 \,\mu}{R} 
\int \frac{\text{d}^4 \ell}{(2\,\pi)^4}
 \sum_{n_1, n_2}
&\bigg[
\frac{n_1 + \alpha + \psi_{n_2}(0)(k_{n_2} + \alpha)}
{
\left[(n_1+\alpha)^2 + \ell^2\right]
\left[(k_{n_2}+\alpha)^2 + \ell^2\right]
}  + \bigg( \textrm{folded sector} \bigg) \bigg].
\label{eq:app_Z_bmu}
\end{align}
The folded sector contribution is obtained by substituting $\alpha \rightarrow \alpha - 1/2$ in that of the MSSM. The right panel of Fig.~\ref{fig:app_Zfactor_deltam} shows the fractional change of the $b_\mu$-term coming from the top sector at $\alpha=1/4$. Again, for comparable $Z'_U \simeq Z'_{U_f} > 0$, the change in $b_\mu$ is around $\lesssim $ 20\%. The $\mathbb{Z}_2^f$ symmetry at $y=0$ will result in correction of order $\mathcal{O}(10\%)$ similar to the Higgs soft-mass.

This highlights the challenge to calculability posed by brane-localized kinetic terms when the 5D cutoff is not parametrically separated from the inverse radius of compactification.  Nevertheless, we conclude that our results are valid across a wide range of parameter space.


\end{spacing}
\begin{spacing}{1.1}
\bibliography{TwistedFoldedSUSY}
\bibliographystyle{utphys}
\end{spacing}
\end{document}